\documentclass[journal]{IEEEtran}
\usepackage[utf8]{inputenc}
\usepackage[T1]{fontenc}

\usepackage{graphicx}
\usepackage{subfig}
\usepackage{float}

\usepackage{amsmath,amssymb,amsfonts}
\usepackage{algpseudocode,algorithm,algorithmicx,float} 

\usepackage{tabularx,booktabs}
\usepackage{multirow}
\usepackage{makecell}
\setcellgapes{3pt}

\usepackage{lipsum}
\usepackage{color, xcolor}

\newcommand{\xmark}{\ding{55}}

\usepackage{caption}
\usepackage[noadjust]{cite}
\usepackage{pifont}
\usepackage{enumitem}
\usepackage{csquotes}

 \def \revision{\color{black}}
  \def \revisiontwo{\color{black}}
 
\makeatletter
\def\ps@IEEEtitlepagestyle{%
  \def\@oddfoot{\mycopyrightnotice}%
  \def\@evenfoot{}%
}
\def\mycopyrightnotice{%
  {\footnotesize  This paper has been accepted for publication by the IEEE Internet of Things Journal. The copyright is with IEEE and the final version will be published by IEEE. \hfill}% <--- Change here
  \gdef\mycopyrightnotice{}% just in case
}

\begin{document}
\bstctlcite{IEEEexample:BSTcontfrol}

\title{CCIC-WSN: An Architecture for Single Channel Cluster-based Information-Centric Wireless Sensor Networks}

\author{
	Muhammad Atif Ur Rehman,
	Rehmat Ullah,
	Byung-Seo Kim,
	Boubakr Nour,
	and Spyridon Mastorakis
	
	\thanks{
	    This work was supported by the National Research Foundation of Korea(NRF) grant funded by the Korea government (No. 2018R1A2B6002399). Corresponding author: Byung-Seo Kim (email:jsnbs@hongik.ac.kr)
	    
		M. A. Ur Rehman is with Department of Electronics and Computer Engineering, Hongik University, Sejong 30016, Republic of Korea.
		
		R. Ullah is with Department of Computer Engineering, Gachon University, Seongnam 13120, South Korea.
		
		B. S. Kim is with Department of Software and Communications Engineering, Hongik University, Sejong 30016, Republic of Korea.
		
		B. Nour is with the School of Computer Science and Technology, Beijing Institute of Technology, Beijing, China.
		
		S. Mastorakis is with the Department of Computer Science, University of Nebraska at Omaha, Omaha, USA
	}	
}

\maketitle

\begin{abstract}
	The promising vision of Information-Centric Networking (ICN) and of its realization, Named Data Networking (NDN), has attracted extensive {\revision attention in recent years} in the context of the Internet of Things (IoT) and Wireless Sensor Networks (WSNs). However, a comprehensive NDN/ICN-based architectural design for WSNs, including specially tailored naming schemes and forwarding mechanisms, has yet to be explored. In this paper, we present single-Channel Cluster-based Information-Centric WSN (CCIC-WSN), an NDN/ICN-based framework to fulfill the requirements of cluster-based WSNs, such as communication between child nodes and cluster heads, association of new child nodes with cluster heads, discovery of the namespace of newly associated nodes, and child node mobility. Through an extensive simulation study, we demonstrate that CCIC-WSN achieves 71–90\% lower energy consumption and 74–96\% lower data retrieval delays than recently proposed frameworks for NDN/ICN-based WSNs under various evaluation settings.
\end{abstract}


\IEEEpeerreviewmaketitle

\section{Introduction}
 \IEEEPARstart{W}{}{\revision ireless Sensor Networks (WSNs) consist of resource-constrained and resourceful nodes in terms of computation power, storage capacity, battery source, and communication range ~\cite{yick2008wireless}. These nodes are often deployed in highly dynamic and sometimes inaccessible areas to perform tasks such as (1) monitoring and sensing, (2) storing of acquired information reports, and (3) transmission of such reports toward a sink node, either directly or via multiple hops. As the acquired information from a sensor node must be forwarded to the sink node for further analysis, efficient forwarding plays an important role in minimizing energy consumption. Limited by communication range and battery life, the direct communication in WSN is not practical as high energy consumption during transmission results in early expiration of a sensor node~\cite{clustersurvey}. At the same time, a significant amount of energy is also consumed in multi-hop communication as the intermediate nodes retransmit the received packets on behalf of neighboring nodes.}

\begin{figure*}
		\centering
		\includegraphics[width=0.9\linewidth]{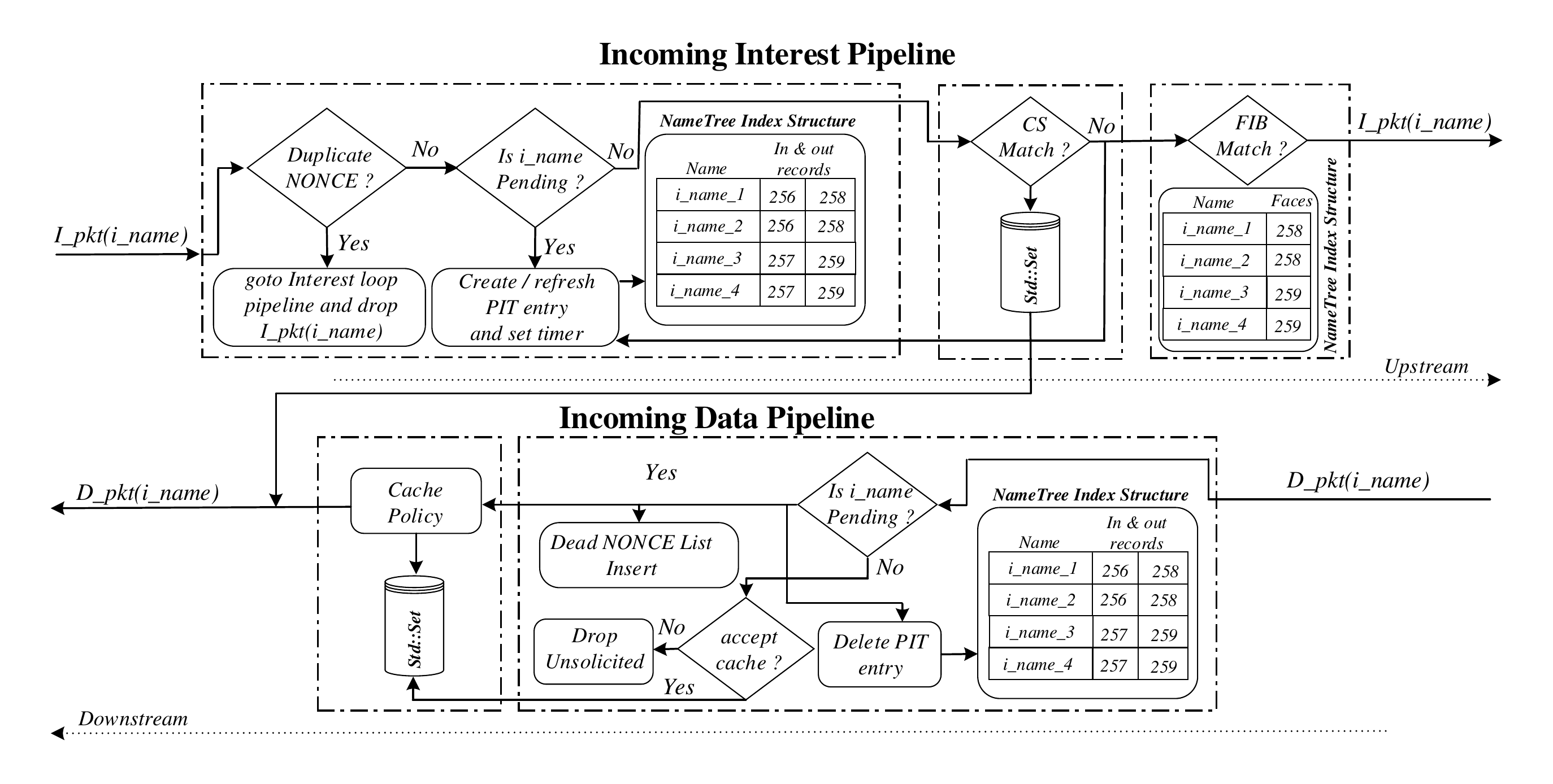}
		\caption{{\revision NDN communication process.}}
		\label{fig2:NDN}
	\end{figure*}
{\revision To address these limitations, a cluster-based hierarchical approach is usually employed in WSN, where the sensor nodes organize themselves into multiple groups ~\cite{cluster-2,cluster-3,salah}. A leader node, often referred to as Cluster Head (CH), is selected and is responsible for performing tasks such as 1) accepting association request from a node that wishes to join the cluster, 2) aggregating the information acquired from child nodes (sensor nodes members of a cluster), and 3) forwarding the aggregated information towards a sink node. In such an architecture, a child node has to forward the sensed information toward the CH located at a single hop and relatively in close proximity as compared to the sink node. As a result, a significant amount of energy can be preserved in a cluster-based WSN.}

{\revision In both cluster-based WSN and general WSN, the end users, such as sink nodes or base stations, are interested in fetching the updated sensed information regardless of which sensor node is producing the information~\cite{nour2019survey}.} In the conventional WSNs, the nodes communicate with each other via an address-based architecture that often introduces several challenges, especially in highly dynamic environments. {\revision Therefore, the recent paradigm shift in communication models from address-oriented to content-oriented} has motivated researchers to look into content-centric network architectures such as Named Data Networking (NDN) ~\cite{NDN}, which employs names rather than host addresses for communication purposes.

{\revision As an architecture, NDN was originally utilized for communication between devices on the Internet. However, the Maximum Transmission Unit (MTU) size and the general resource requirements of these devices are different from those in WSNs. Such characteristics result in the need} to extend the existing NDN architecture for communication in WSNs. For instance, the namespace design for WSNs should consider a rather limited length of names that can fit into 127 bytes MTU~\cite{ISI}. Moreover, for cluster-based WSNs, separate naming schemes should be designed for CHs and child nodes, as these nodes may vary in terms of the available resources. {\revision Considering that WSNs follow a wireless ad hoc communication model, energy-efficient forwarding mechanisms are also required to minimize unnecessary packet transmissions.} 

{\revision Taking into account all these unique characteristics of cluster-based WSNs,} this paper proposes extensions to the NDN architecture for WSNs. We first propose naming schemes for {\revision CHs and child nodes}. We also propose a naming scheme for the association of new nodes with CHs, which is an essential process in cluster-based WSNs, and the sharing of CH information with new, not-yet-associated nodes. After the association of a new node, a CH employs a synchronization process to share the name of the newly associated node with other nodes in the network for communication purposes. Our proposed architecture, coined single Channel Cluster-based Information-Centric Wireless Sensor Networks (CCIC-WSN), supports child node mobility by having CHs maintain a data structure with the members of their clusters (i.e., the child nodes that are associated with them) {\revision and} synchronize this data structure with other CHs. Finally, CCIC-WSN features energy-efficient forwarding mechanisms for intra- and inter-cluster communication.

{\revision The main contributions of our work are summarized as follows:}

\begin{itemize}[leftmargin=*]
	\item \textbf{Naming schemes for WSNs:} {\revision CCIC-WSN offers a set of naming schemes for: (i) the exchange of data among resource-rich CHs and resource-constrained child nodes; and (ii) the association of new nodes with CHs.}
	
	\item \textbf{Lite-query structure:} {\revision CCIC-WSN supports the execution of named lite queries on CHs for efficient data extraction, with the help of different constraints. }
	
	\item \textbf{Named-based forwarding mechanisms for WSNs:} CCIC-WSN features a set of name-based forwarding mechanisms for {\revision energy-efficient} intra- and inter-cluster (over multiple wireless hops) communication, minimizing unnecessary packet transmissions.

	\end{itemize}	
	
Through our extensive simulation study, we demonstrated that CCIC-WSN significantly outperforms recently proposed NDN-based frameworks for WSNs, achieving 71–90\% lower energy consumption and 74–96\% lower data retrieval delays than these frameworks. To the best of our knowledge, CCIC-WSN is the first comprehensive architectural design for cluster-based WSNs in NDN. 

The rest of this paper is organized as follows: Section~\ref{sec:background} provides some brief background on NDN and some related work{\revision,} while Section~\ref{sec:motivation} motivates the design of CCIC-WSN. {\revisiontwo An architectural description of the CCIC-WSN design is outlined in Section~\ref{sec:architectureovewview}.   Section~\ref{sec:components} specifies the details of various components of CCIC-WSN}.  Section~\ref{sec:evaluation} presents our simulation study. Subsequently, a number of open research challenges and future directions are discussed in Section~\ref{sec:future}{\revision; Section~\ref{sec:conclustion} finally concludes our work.}

\begin{table*}[!t]
	\centering
% 	\makegapedcells
	\caption{Summary of Related Work}
	\begin{tabular}{|c|c|c|c|c|c|c|}
		\hline
		\textbf{References}                       & \multicolumn{1}{m{5cm}|}{\textbf{Objective}}     & \textbf{Query Support}& \textbf{Pull Support} & \textbf{Push Support}& \textbf{Mobility}& \textbf{Clustering}                                                                 \\ \hline
		
		\cite{UWSN}                       & \multicolumn{1}{m{5cm}|}{An NDN system design for under water WSN}      & \xmark & \checkmark & \xmark& \xmark & \xmark                                                                                                        \\ \hline
		
		\cite{HFHN}                       & \multicolumn{1}{m{5cm}|}{An ICN-based pull- \& push-based communication framework for smart buildings}      & \xmark & \checkmark & \checkmark& \checkmark & \xmark                                                                  \\ \hline
		
		\cite{NDWSN}    & \multicolumn{1}{m{5cm}|}{An NDN system design for data collection in WSN }      & \xmark & \checkmark & \xmark& \xmark & \xmark                                                                                                        \\ \hline
		
		\cite{PBD2} & \multicolumn{1}{m{5cm}|}{Push data broadcast control in ICN multi-hop wireless networks}      & \xmark & \checkmark & \checkmark& \checkmark & \xmark                                                                                                        \\ \hline
		
		\cite{NINQ}                       & \multicolumn{1}{m{5cm}|}{A name integrated query framework for NDN-based IoT}      & \checkmark & \checkmark & \checkmark& \checkmark & \xmark                                                                            \\ \hline
		
		\cite{CCN-WSN}                       & \multicolumn{1}{m{5cm}|}{A lightweight, flexible Content-Centric
			Networking Protocol for WSN}      & \checkmark & \checkmark & \checkmark& \xmark & \xmark                                                 \\ \hline
		
		\cite{WR}                       & \multicolumn{1}{m{5cm}|}{An NDN based real time wireless recharging
			framework for WSN}      & \xmark & \checkmark & \xmark & \xmark & \xmark                                                                   \\ \hline
		
		\cite{EEIF}                       & \multicolumn{1}{m{5cm}|}{An energy efficient Interest forwarding mechanism in WSNs}      & \xmark & \checkmark & \checkmark& \checkmark & \xmark                                                                                             \\ \hline
		
		CCIC-WSN                       & \multicolumn{1}{m{5cm}|}{An ICN architectural design for cluster-based WSNs}      & \checkmark & \checkmark & \checkmark& \checkmark & \checkmark                                                                                          \\ \hline
	\end{tabular}%
	\label{table:related}
	\vspace{-0.2cm}
\end{table*}

\section{Background \& Related Work}
\label{sec:background}
\subsection {{\revision Named Data Networking: An overview}}
\label{subsec:ndnoverview}
NDN~\cite{NDN} is a realization of the Information-Centric Networking (ICN) vision~\cite{xylomenos2013survey}. In NDN/ICN, the communication model is based on the content name. Consumer applications send requests for named content/data, {\revision called \emph{Interest packets} toward producer} applications that generate the requested content. In response to Interests, producers send the requested content, in the form of a \emph{Data packet}, back to the requesting consumer(s). Once generated, Data packets are cryptographically signed by their producers, binding the generated content to the name of the Data packets that carry this content.

NDN is based on the following fundamental principles: (i) \textbf{application-defined, hierarchical, and semantically meaningful naming:} NDN carries packets that are identified through application-defined names directly at the network layer for communication purposes; (ii) \textbf{name-based, stateful forwarding:} NDN forwards Interests based on their {\revision names toward the requested data---forwarded Interests leave state at each NDN}  forwarder~\cite{nour2020collaborative}, while the retrieved Data packets utilize this state to follow the same path as their corresponding Interests back to the requesting consumer(s), satisfying the state at each NDN forwarder~\cite{mastorakis2019towards, mastorakis2020icedge}; and (iii) \textbf{content-centric security:} each Data packet in NDN carries the signature of its producer directly at the network layer, which secures the content in transit across the network and at rest~\cite{zhang2018overview}.

{\revision A detailed Interest/Data packets forwarding process in NDN is illustrated in Fig.~\ref{fig2:NDN} and briefly discussed below.} To realize the NDN stateful forwarding plane, NDN forwarders are equipped with three data structures: (i) a Forwarding Information Base (FIB): FIB contains a number of entries, where each entry consists of a name prefix and a set of outgoing interfaces, and is used for Interest forwarding purposes; (ii) a Pending Interest Table (PIT): PIT stores recently forwarded Interests that have yet to bring content back, essentially maintaining the network state created by these Interests; and (iii) a Content Store (CS): CS caches recently retrieved Data packets to satisfy future requests for the same content.
\begin{figure*}
		\centering
		\includegraphics[width=0.9\linewidth]{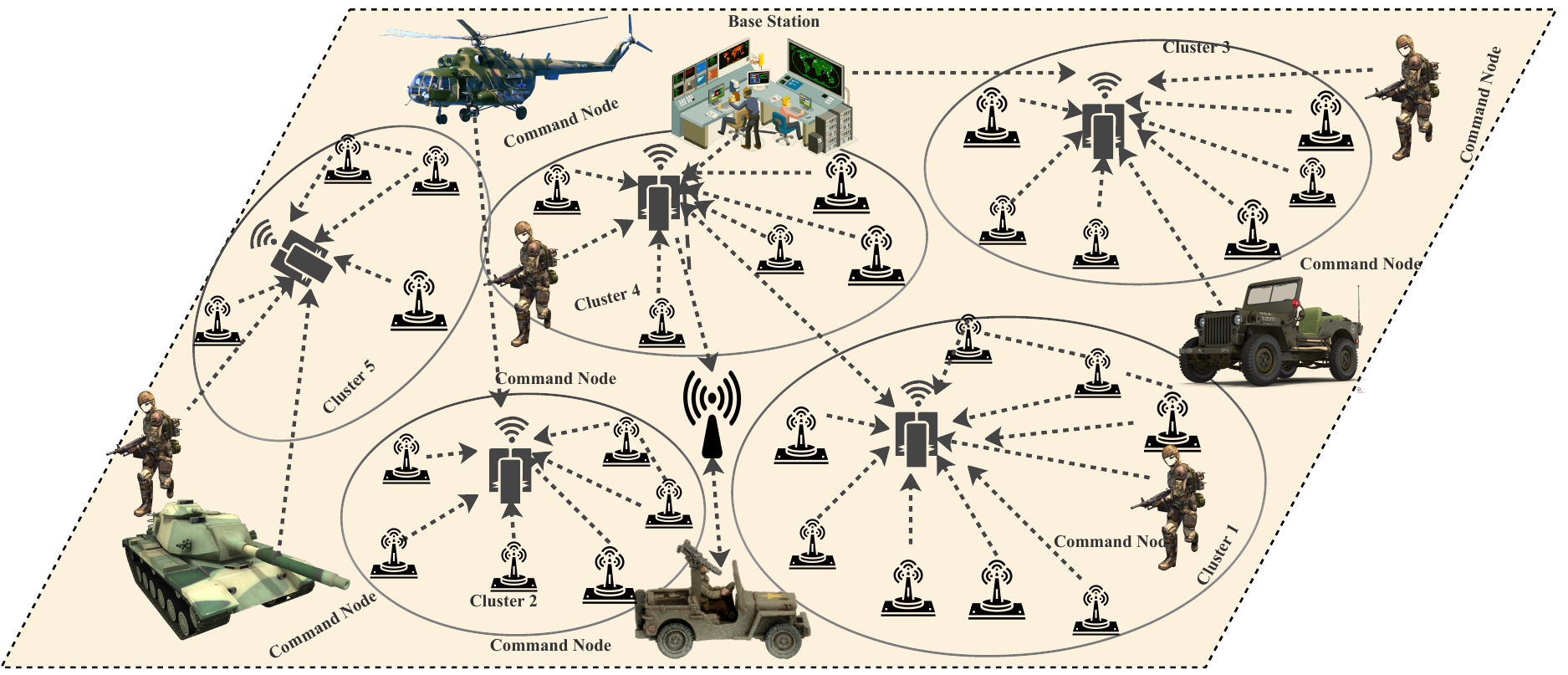}
		\caption{{\revision Cluster-based WSN for Battlefield Surveillance.}}
		\label{fig:motivation}
\end{figure*}

\subsection {{\revision Related Work: WSNs over NDN/ICN}}
\label{subsec:relWSN}
{\revision Table~\ref{table:related} presents a summary of related work on WSNs over NDN/ICN}. Bouk {\it et al.}~\cite{UWSN} utilize hierarchical naming for pull-based communications in underwater sensor networks, where the selected naming components refer to time, location, data type, and preference. Arshad {\it et al.}~\cite{HFHN} consider the scenario of smart buildings and introduce a hybrid naming scheme consisting of hierarchical and flat components, without specifically considering the deployment of cluster-based wireless nodes in smart buildings. Furthermore, Rehman {\it et al.}~\cite{CCIC} and Amadeo {\it et al.}~\cite{NDWSN} perform preliminary explorations of the design space, propose directions for the integration of WSN with NDN, without proposing complete architectural designs. Ullah {\it et al.}~\cite{PBD2},~\cite{ullah2019push} consider push-based use-cases (e.g., accident reporting for vehicular networks) {\revision and introduce a} naming scheme to control the propagation of data packets in the network. To mitigate potential storms of broadcast messages, {\revision the authors propose modifications} of the overall NDN forwarding and packet processing logic. Furthermore, Rehman {\it et al.} propose a flexible, expressive, and secure framework that offers a query-based mechanism for content retrieval in smart building scenarios~\cite{NINQ, rehman2019network}. This framework also supports configuration and control commands among nodes in a smart building. However, the query structure is not compatible with the 127-byte MTU size of WSNs, {\revision while neither of the two solutions considered functions such as  new node association that are needed in cluster-based WSNs.} 

To accomplish time synchronization in ICN-based WSN networks, a protocol was proposed where clock information of nodes is shared via ICN messaging~\cite{CCN-WSN}. The authors proposed a modified format of Interest and Data packets to fulfil the requirements of WSNs such as in-network processing and data aggregation. A framework for the real-time recharging of WSN nodes over NDN was also proposed~\cite{WR}. In this work, NDN-based protocols for energy monitoring and reporting were designed {\revision and an analytical result was collected. In addition, an Interest forwarding scheme} for WSNs with two modes of operation (flooding and directive forwarding modes) was proposed~\cite{EEIF}. Data producers are initially discovered via the Interest flooding mode, while directive mode is subsequently used to steer future {\revision Interests toward those producers}. Potential message broadcast storms in the network are mitigated with the use of scope control and packet suppression algorithms~\cite{nour2018nncp}.

% \noindent
\textbf{How does CCIC-WSN differ from prior work?} {\revision The non-cluster-based information-centric WSN approaches discussed thus far primarily deal with the Interest and Data packet forwarding, providing various solutions such as scope control, packet suppression, and deferred timers to avoid broadcast storms, among others. Going beyond the forwarding strategies, the cluster-based information-centric WSNs also require: (i) novel naming schemes that ought to demonstrate good compatibility with resource-constrained and resource-rich nodes for content dissemination; (ii) tailored named and integrated computing functionalities for the CH node that can aggregate the acquired information before data transmission to a sink node; (iii) an efficient mechanism that aims to provide the association of new nodes in the network; (iv) support for child node mobility among various clusters; and (v) a specially designed forwarding process for intra- and inter-cluster communication. The previous works in the domain of information-centric WSN do not provide the necessary mechanisms to deal with these requirements of a cluster-based network. To this end, the proposed CCIC-WSN framework extends the NDN architecture in order to provide efficient communication while supporting the fundamental requirements of cluster-based systems.}

\begin{figure*}
	\centering
	\includegraphics[width=0.9\linewidth]{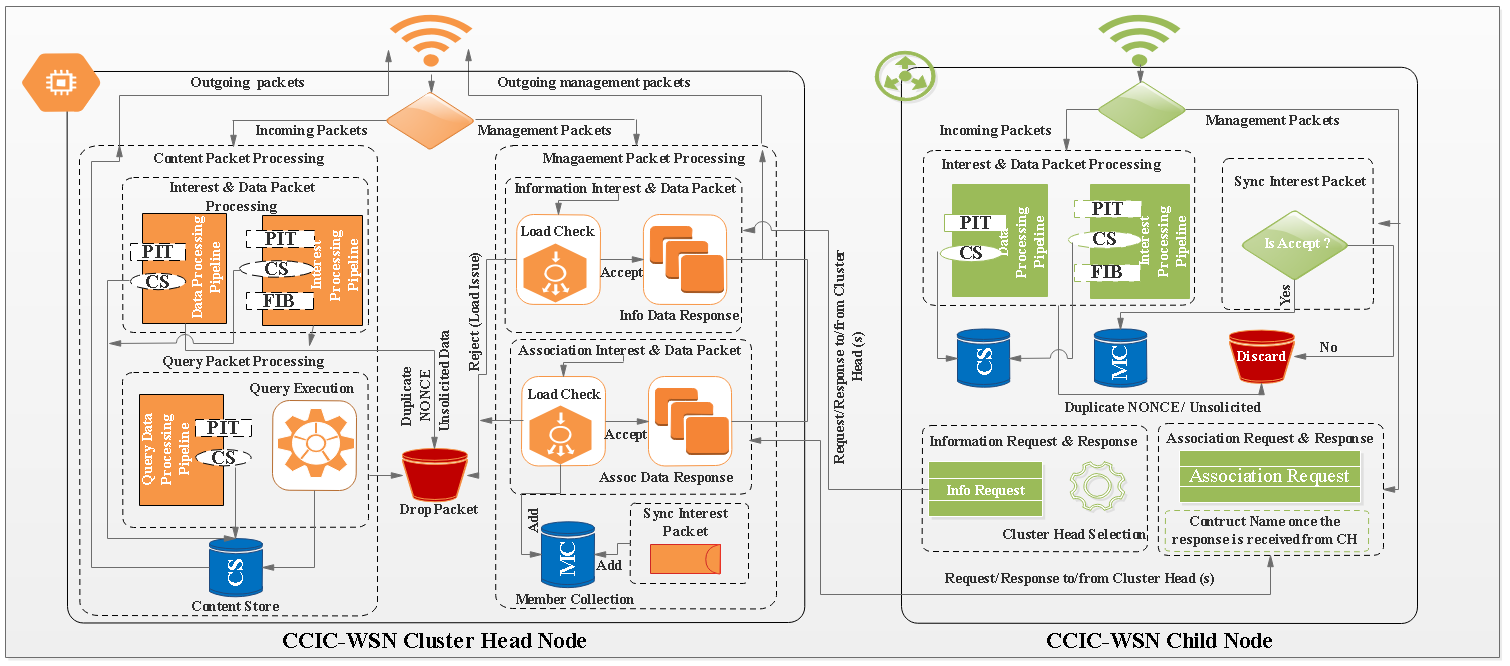}
	\caption{{\revision CCIC-WSN Node Architecture.}}
	\label{fig:CCIC}
\end{figure*}

\section{Motivation}
\label{sec:motivation}
{\revision To motivate the design of CCIC-WSN, let us consider a battlefield surveillance and monitoring system use-case with the help of a diagram presented in Fig.~\ref{fig:motivation}. Such a system may require multiple nodes that perform functions such as collaborative target tracking, sensitive area monitoring, data analysis, and communication. These nodes can be deployed in the form of clusters on the battleground and the nearby suspicious areas. The battlefield surveillance system can be utilized for the monitoring of various objects such as spying robots, tanks, and the movement pattern of enemy troops. Furthermore, this system generates alerts and forwards the acquired information towards base stations in addition to nearby friendly troops to take precautionary measures and immediate response.

Such a system may consist of three layers for: (i) sensing and monitoring; (ii) data aggregation and analysis; and (iii) data fusion. The nodes in these layers may be loosely or strongly coupled with each other for efficient monitoring, target tracking, analysis, and generation of alerts. For example, nodes such as MICAz motes~\cite{micaz} in a sensing and monitoring layer measure acoustic signals produced by various targeted objects. These nodes, often termed as child nodes, are responsible for performing sensing and monitoring operations. The nodes in the second layer such as cluster-heads---capable of performing various compute and storage-intensive tasks~\cite{krol2019compute}---may periodically fetch, aggregate, and examine the recorded samples. Finally, the aggregated information is forwarded towards the base station for data fusion and further analysis.} 

The current deployment of {\revision WSNs, as in the use-case} mentioned above, employ traditional address-based solutions which may result in efficiency and scalability issues. {\revision For example, extracting vibration data from a single sensor in the battlefield surveillance and monitoring system requires the knowledge of the address of each sensor as well as the type of data each sensor generates (address to content/data mapping).} As a result, the management of all nodes in large-scale environments (e.g., enterprises) becomes a significantly complex and time-consuming task. {\revision On the other hand, CCIC-WSN runs on top of NDN eliminating the need for address configuration and tedious content-address mappings as well as supporting the fundamental requirements of cluster-based WSNs.}

%\section{CCIC-WSN Design}
\section{{\revisiontwo CCIC-WSN: An Architectural Overview}}
\label{sec:architectureovewview}
{\revisiontwo In this section, we present an overview of the CCIC-WSN design and we discuss its system model along with our assumptions.}

% , but, at the same time, maintains backward compatibility.

\subsection{System Design in a Nutshell}

{\revision CCIC-WSN provides named services to the nodes in cluster-based WSNs by leveraging the NDN communication model and extending the NDN architecture. The overall system design and, more specifically, the CCIC-WSN node architecture for both a CH and a child node is presented in Fig.~\ref{fig:CCIC}. Novel packet types along with the legacy Interest/Data packets as well as new modules are added to enable the cluster-based communication in WSN as briefly described below.

The incoming packets on a CH (left of Fig.~\ref{fig:CCIC}) are classified into two categories: (i) content request/response packets that are further divided into conventional Interest/Data packets and query-based Interest/Data packets; and (ii) management packets. The management packets and the relevant modules serve as a key enabler of cluster-based communication in the WSN and follow the rituals of the NDN communication architecture. To accomplish the requirements of managing the cluster-based communication, we designed a hierarchical naming scheme for the cluster information association process that identifies the CHs in the network for association purposes. Further, the hierarchical synchronization (\enquote{sync} for short) naming scheme and related module are designed to share association-related information with other nodes in the WSN.

Similar to the CH, the incoming packets on a child node (right part of Fig.~\ref{fig:CCIC}) are also classified into two categories: (i) content request/response packets; and (ii) management packets. To associate with a CH, a child node utilizes information request/response packets to fetch the information of potential CHs in the vicinity and association request/response packets to associate itself with a specific CH. Furthermore, a child node may receive a sync-information packet, containing information about new nodes and may store it in a member collection data structure, if enabled.}

\subsection{System Model and Assumptions}
{\revision In this work, we consider a single-channel cluster-based WSN as a graph $G = (N, A)$, where $N$ is a set of nodes and $A$ is a set of links}. Nodes organize themselves in the form of clusters~\cite{clustering}. {\revision Let $x_n$, a Boolean variable, indicate the resource availability at node $n \in N$. If $x_n = 1$, the node has enough resources in terms of computational power, storage capacity, battery, and transceiver power {\revision to become a CH}; otherwise, $x_n = 0$.} Based on the available resources, there could be at least two types of nodes in each cluster: {\revision Cluster Head ($\mathsf{CH}$) and Child Node ($\mathsf{CN}$)}. {\revision $\mathsf{CH}$} nodes often have more resources compared to {\revision $\mathsf{CNs}$} nodes. {\revision $\mathsf{CH}$} nodes perform cluster management tasks such as the assignment of duty-cycles to {\revision $\mathsf{CNs}$}, reception of sensed data from {\revision $\mathsf{CNs}$}, aggregation operations on received data, and forwarding of aggregated data to the sink or the requesting node. In contrast, the {\revision $\mathsf{CNs}$} have limited resources; therefore, they only sense the environment and {\revision forward sensed data toward a {\revision $\mathsf{CH}$} node located one hop away. }
\begin{table}[!t]
	\caption{Packet Types in CCIC-WSN}
	\begin{tabular}{|c|c|}
		\hline
		\textbf{Packet Type}                       & \textbf{Description}                                                                                                        \\ \hline
		
		\multicolumn{1}{|m{3cm}|}{Cluster Head Interest \& Data Packets} & 
		\multicolumn{1}{m{5cm}|}{These conventional Interest \& Data packets are used to fetch the content from a cluster head} 
		\\ \hline
		\multicolumn{1}{|m{3cm}|}{Child Node Interest \& Data Packets} & 
		\multicolumn{1}{m{5cm}|}{These conventional Interest \& Data packets are used to fetch the content from a child node} 
		\\ \hline
		\multicolumn{1}{|m{3cm}|}{Cluster Head Information Interest Packet} & 
		\multicolumn{1}{m{5cm}|}{An unassociated node forwards this packet to find the potential CHs in its neighborhood.} 
		\\ \hline
		\multicolumn{1}{|m{3cm}|}{Cluster Head Information Data Packet} & 
		\multicolumn{1}{m{5cm}|}{A CH forwards this response after receiving a CH information Interest packet} 
		\\ \hline
		\multicolumn{1}{|m{3cm}|}{Cluster Head Association Interest Packet} & 
		\multicolumn{1}{m{5cm}|}{An unassociated node forwards this packet to associate itself with a CH} 
		\\ \hline
		\multicolumn{1}{|m{3cm}|}{Cluster Head Association Data Packet} & 
		\multicolumn{1}{m{5cm}|}{A CH acknowledges the cluster head association Interest of a child node} 
		\\ \hline
		\multicolumn{1}{|m{3cm}|}{New Node Sync Interest Packet} & 
		\multicolumn{1}{m{5cm}|}{This packet is used to share the information of newly associated child node}
		\\ \hline
		\multicolumn{1}{|m{3cm}|}{New Node Sync Data Packet} & 
		\multicolumn{1}{m{5cm}|}{This packet acknowledges the reception of a sync Interest packet} 
		\\ \hline
	\end{tabular}%
	\label{Table:packettypes}
\end{table}

In cluster-based WSN, {\revision both $\mathsf{CH}$ and $\mathsf{CNs}$} can act as producers of data. At the micro level, the {\revision $\mathsf{CNs}$} are deployed at specific locations and sense their surroundings. At this level, a {\revision $\mathsf{CN}$} acts as a producer of data, while a {\revision $\mathsf{CH}$} node acts as a consumer.  
{\revision Considering that the $\mathsf{CN}$ acts as a provider,} a specific naming scheme is required that assigns a unique and compact name to the sensed data. {\revision Eq.~\ref{eq:name_cn} {\revision expresses} the assignment of a naming scheme (type 1) for the sensed data by a $\mathsf{CN}$.} 

\begin{equation} \label{eq:name_cn}
    \mathsf{CN} \leftarrow \text{Naming}_{1} . S^{d}
\end{equation}

{\revision As opposed to this,} at the macro level, a {\revision $\mathsf{CH}$} node acts as a producer of data generated by several {\revision $\mathsf{CNs}$} under its administration. {\revision As a {\revision $\mathsf{CH}$} is responsible for several {\revision $\mathsf{CNs}$}, which may be heterogeneous in terms of} sensing the environment (e.g., audio, video, or scalar sensors), the naming scheme of {\revision $\mathsf{CH}$} nodes differs from the naming scheme of {\revision $\mathsf{CNs}$}. {\revision Eq.~\ref{eq:name_ch} expresses the  assignment of a naming scheme (type 2) for the generated data by a $\mathsf{CH}$.} 

\begin{equation} \label{eq:name_ch}
    \mathsf{CH} \leftarrow \text{Naming}_{2} . G^{d}
\end{equation}

Given that asymmetric cryptography may be too expensive for a WSN use-case~\cite{shi2004designing}, we assume that {\revision $\mathsf{CN}$} and {\revision $\mathsf{CH}$} nodes are able to perform symmetric cryptographic algorithms. To this end, we assume that nodes can utilize algorithms such as Advanced Encryption Standard (AES)~\cite{daemen2013design} and Hashed Message Authentication Codes (HMAC)~\cite{krawczyk1997hmac} to protect the integrity and authenticate the generated Data packets. We discuss directions for secure node {\revision association and data encryption/decryption} and authentication in Section~\ref{subsec:authentication}.
Table~\ref{Table:packettypes} shows an overview of the different types of Interest and Data packets that are used in CCIC-WSN. In the following subsections, we provide a detailed description of all of these packet types and the components of the CCIC-WSN design.
%{\todo \bf I think it will be better to add a dedicated subsection regarding the architecture design and components, where we explain why we need to add more packets (Table II)!}

\section{{\revisiontwo CCIC-WSN Design Components}}
\label{sec:components}
{\revisiontwo In this section, we present the components of the CCIC-WSN design in detail.}

\subsection{Naming for content fetching from {\revision $\mathsf{CH}$} nodes and {\revision $\mathsf{CNs}$}}

\begin{figure}[!t]
	\centering
	\includegraphics[scale=0.45]{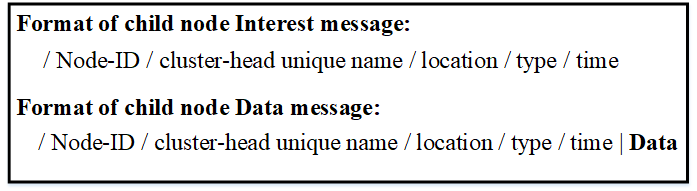}
	\caption{{\revision $\mathsf{CN}$} node namespace design.}
	\label{fig:cnname}
\end{figure}

\textbf{{\revision $\mathsf{\textbf{CN}}$ Namespace Design:}}
The namespace design for {\revision $\mathsf{CN}$, as expressed in Eq.~\ref{eq:name_cn}}, is presented in Fig.~\ref{fig:cnname}. This namespace consists of several components: 
The first component refers to the {\revision $\mathsf{CN}$ ID, which is used during the new node association process and} by {\revision $\mathsf{CHs}$} to differentiate the data produced by different {\revision $\mathsf{CNs}$} under their administration; {\revision the} second component refers to the name of the {\revision $\mathsf{CH}$} and is used during intra-cluster forwarding and to limit the number of redundant packet transmissions; {\revision the} third component refers to the physical location of a {\revision $\mathsf{CN}$}, which can be obtained through a Global Positioning System (GPS) module installed on this node; {\revision the} fourth component refers to the type of the data sensed (produced) by the node (e.g., temperature, pressure, audio, video{\revision , etc.}); and the last component refers to the time at which the sensed information was produced\footnote{Note that, for the CCIN-WSN design, we chose the epoch time format to save extra bytes required by a general time, where the time is separated through different notations, such as “:”, “-“ and “/”.}.

\begin{figure}[!t]
	\centering
	\includegraphics[scale=0.4]{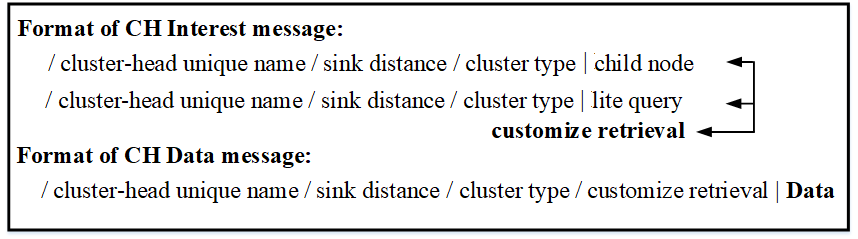}
	\caption{{\revision $\mathsf{CH}$} namespace design.}
	\label{fig:chname}
\end{figure}

\textbf{{\revision $\mathsf{\textbf{CH}}$ Namespace Design:}}
A {\revision $\mathsf{CH}$} acts as a consumer when it fetches content from a {\revision $\mathsf{CN}$} {\revision (c.f., Eq.~\ref{eq:name_cn})} and as a producer when other {\revision $\mathsf{CH}$}s or sink nodes fetch the content from it {\revision (c.f., Eq.~\ref{eq:name_ch} )}. To this end, a {\revision $\mathsf{CH}$} should have a proper namespace that incorporates the type of content that is produced under its administration. Fig.~\ref{fig:chname} illustrates the namespace design for {\revision $\mathsf{CH}$} nodes, which includes several components: 
The first component refers to the {\revision $\mathsf{CH}$} name prefix, which is used to locate the {\revision $\mathsf{CH}$} in multi-hop communication scenarios; 
{\revision the} second one specifies the distance of the {\revision $\mathsf{CH}$} to the sink node; 
{\revision the} third one is determined based on the the responsibilities of the {\revision $\mathsf{CN}$} (type of generated data). The {\revision $\mathsf{CH}$} selects the cluster type at the time of network initialization. To accept the association requests from various types of {\revision $\mathsf{CNs}$}, the {\revision $\mathsf{CH}$} selects "heterogeneous" as the cluster type. Alternatively, the {\revision $\mathsf{CH}$} selects a specific cluster type (e.g., temperature, scalar, audio, video{\revision , etc.}), so that only {\revision $\mathsf{CNs}$} that generate the specified type of data can be associated with {\revision this {\revision $\mathsf{CH}$}. The} last component may have two types of {\revision values: It may either contain a lite-query} (Section~\ref{subsec:lite}) or it can specify the name of a {\revision $\mathsf{CN}$} to fetch data from that {\revision $\mathsf{CN}$}.

\subsection{Customized Lite-Query Structure for WSN}
\label{subsec:lite}
The lite-query component is the last part of the {\revision $\mathsf{CH}$} naming scheme and can only be processed by the {\revision $\mathsf{CH}$} node. We limit the {\revision processing of lite queries to the {\revision $\mathsf{CH}$} because it} usually has enough computing, storage, and battery resources, which are required for the processing and storing of the results that a query may yield. We present examples of various lite-queries that we use in CCIC-WSN in Table~\ref{table:lite}. Note that the query scope is not limited to these examples; many other combinations based on the WSN application requirements can be defined. Each component in the lite-query is separated by “\_” and plays an important role in content filtering. The first component in the lite-query is a combination of two dynamic keywords, where a character “.” is used to separate these two keywords. The first keyword denotes the task name{\revision ,} such as vibration, temperature, pressure, while the second keyword defines the name of the field{\revision ,} such as date, time, NID (Node ID), or val (value). 

\begin{table}[!t]
	\caption{Interpretation of Lite-Queries in CCIC-WSN }
	\begin{tabular}{|c|c|}
		\hline
		\textbf{Lite-Query Example}                       & \textbf{Description} \\ \hline

		\multicolumn{1}{|m{3cm}|}{tem.val\_gt\_25} & 
		\multicolumn{1}{m{5cm}|}{Select a temperature sub-collection in which values are greater than 25} 
		\\ \hline
		\multicolumn{1}{|m{3cm}|}{tem.val\_lt\_25} & 
		\multicolumn{1}{m{5cm}|}{Select a temperature sub-collection in which values are less than 25} 
		\\ \hline
		\multicolumn{1}{|m{3cm}|}{tem.val\_eq\_25} & 
		\multicolumn{1}{m{5cm}|}{Select a temperature sub-collection in which values are equal to 25} 
		\\ \hline
		\multicolumn{1}{|m{3cm}|}{tem.val\_neq\_25} & 
		\multicolumn{1}{m{5cm}|}{Select a temperature sub-collection in which values are not equal to 25} 
		\\ \hline
		\multicolumn{1}{|m{3cm}|}{tem.val\_in\_25} & 
		\multicolumn{1}{m{5cm}|}{Check whether the temperature collection has a value of 25 or not. The result will contain a Boolean value} 
		\\ \hline
		\multicolumn{1}{|m{3cm}|}{tem.val\_bet\_25\_and\_50} & 
		\multicolumn{1}{m{5cm}|}{Select a temperature sub-collection in which values are between 25 and 50} 
		\\ \hline
		\multicolumn{1}{|m{3cm}|}{tem.val\_gt\_25\_limit\_10\_dsc} & 
		\multicolumn{1}{m{5cm}|}{Select the top 10 (in a descending order) temperature sub-collections in which values are greater than 25} 
		\\ \hline
		\multicolumn{1}{|m{3cm}|}{tem.val\_gt\_25\_count} & 
		\multicolumn{1}{m{5cm}|}{Retrieve the number of temperature sub-collections in which values are greater than 25} 
		\\ \hline
		\multicolumn{1}{|m{3cm}|}{tem.val\_bet\_9\_and\_14\_avg} & 
		\multicolumn{1}{m{5cm}|}{Get the average temperature values of the current day between 9AM and 2PM} 
		\\ \hline
	\end{tabular}%
	\label{table:lite}
\end{table}

To compress the size of the lite-query and save 5--10 bytes in the content name, we limit the size of the task name to three characters. For example, a task “temperature” {\revision can be defined through “tem,”} while a task “vibration” can be defined through “vib”. Therefore, by using {\revision 3} characters from 52 alphabets (small and large caps), our structure can scale up to approximately 52\textsuperscript{3} unique tasks. 
\begin{figure}[!t]
	\centering
	\includegraphics[scale=0.40]{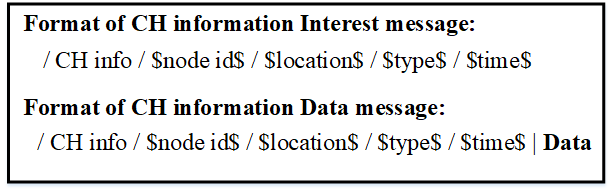}
	\caption{CH selection namespace.}
	\label{fig:chsname}
\end{figure}
The second keyword that refers to the field name can be a maximum of 4 bytes. These {\revision two keywords are combined} and form a unique dynamic keyword that is employed to filter the content. For instance, “vib.time” indicates that the requested vibration data should be filtered based on the time value of this lite-query. The second component in the lite-query structure is the comparison operator. For example, the comparison operator “lte” (less than or equal to) can be employed to compare the values of \texttt{<task>.<field>} with a static value that follows this operator. Several other keywords such as desc, asc, count, sel, avg, min, and max can be used to filter the results. 

\subsection{{\revision $\mathsf{\textit{CH}}$} Join Process}
When joining a network, {\revision a new (unassociated) node} selects a {\revision $\mathsf{CH}$} to become part of a network. In CCIC-WSN, a {\revision $\mathsf{CH}$} responds to cluster selection and {\revision $\mathsf{CH}$} association requests. The {\revision $\mathsf{CH}$} join process consists of two phases: i) the {\revision $\mathsf{CH}$} {\revision selection phase} and ii) the {\revision $\mathsf{CH}$} association phase. {\revision As a result, the total time needed for the join process $t_{join}$ is:

\begin{equation} 
     t_{join} = t_{selection} + t_{association}
     \label{eq:join}
\end{equation}

\noindent where $t_{selection}$ and $t_{association}$ represent the time for the $\mathsf{CH}$ selection and the $\mathsf{CH}$ association phases respectively.
}

\subsubsection{{\revision $\mathsf{CH}$} Selection Phase}
In the {\revision $\mathsf{CH}$} selection phase, a new node broadcasts an Interest packet with a name as shown in Fig.~\ref{fig:chsname}, which consists of the following components:

\textit{CH\_Info:} This is a prefix used to indicate that the current Interest is a {\revision $\mathsf{CH}$} selection request. This component is used as a routing prefix to locate multiple {\revision $\mathsf{CH}$}s in the network.

\textit{Node\_ID:} This is the ID of the new node. From the {\revision $\mathsf{CH}$} perspective, this name component is variable and can take various values. The {\revision $\mathsf{CH}$} uses this ID to distinguish the current node from other {\revision CNs} and their produced data as well.

\textit{Location:} These are the physical coordinates of the new node and can be used by a {\revision $\mathsf{CH}$} to decide on whether to accept the current node as a child. 

\textit{Type:} This is the type of the new node based on the data it produces (e.g., temperature, pressure, video, audio {\revision , etc.}). Similar to the location component, a {\revision $\mathsf{CH}$} uses this component to decide whether to accept the current node as a child.

\textit{Time:} It shows that the data from a new node can be accessed for a specific time interval.

The above hierarchical naming format consists of five components. {\revision Among them, only the first component is static and is used to locate the {\revision CHs} in the network; the remaining four components are variable, and their values depend on the unassociated node.} On receiving such an Interest, a {\revision $\mathsf{CH}$} may or may not respond depending on its current workload. If there is no excessive workload on the {\revision $\mathsf{CH}$}, the {\revision $\mathsf{CH}$} will respond with a {\revision Data packet} that carries {\revision $\mathsf{CH}$} related information, such as the {\revision $\mathsf{CH}$} distance from a sink node, the number of {\revision CNs} that are already associated with the {\revision $\mathsf{CH}$}, the traffic load on the {\revision $\mathsf{CH}$}, and the unique name of the {\revision $\mathsf{CH}$} itself for {\revision $\mathsf{CH}$} name discovery. 

\subsubsection{{\revision $\mathsf{\textit{CH}}$} Association Phase}
When a new node receives a selection response from one or more {\revision $\mathsf{CH}$}s, it decides which {\revision $\mathsf{CH}$} to associate with. The {\revision $\mathsf{CH}$} selection decision may be based on various parameters and is out of the scope of our work. After deciding which {\revision $\mathsf{CH}$} to select, a {\revision $\mathsf{CN}$} forwards a {\revision $\mathsf{CH}$} association Interest that carries a name as shown in Fig.~\ref{fig:chassocname}. This name has the following components:

\begin{figure}[!t]
	\centering
	\includegraphics[scale=0.38]{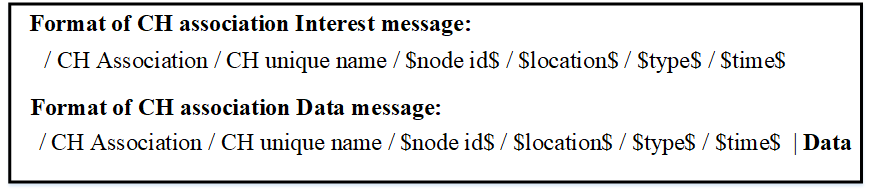}
	\caption{{\revision $\mathsf{CH}$} association namespace.}
	\label{fig:chassocname}
\end{figure}

\textit{CH\_Association:} It indicates that the current Interest packet is an association request {\revision from an unassociated node.} 

\textit{CH Unique Name:} It indicates that the current Interest packet is only destined for the CH whose name is mentioned in this field.  

\begin{figure}[!t]
	\centering
	\includegraphics[scale=0.56]{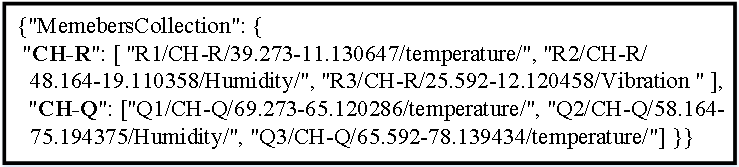}
	\caption{An example of a members collection}
	\label{fig:memcollec}
\end{figure}

The remaining components in the name are the same as presented in Fig.~\ref{fig:chsname}. Once a CH receives an association Interest, it uses the name components of this Interest to create Interests that will fetch data from this {\revision $\mathsf{CN}$} in the future. It further stores the information of the {\revision $\mathsf{CN}$} in its members collection as shown in Fig.~\ref{fig:memcollec}. After storing this information, the CH performs two operations: i) responds with a {\revision Data packet to the  $\mathsf{CN}$} and ii) starts a sync process in the network (further explained in Section~\ref{subsec:sync}). The CH association Data packet acts as an acknowledgment to the {\revision $\mathsf{CN}$} that the CH has accepted its association request. {\revision Upon receiving this Data packet,} the {\revision $\mathsf{CN}$} completes its own name by adding the CH unique name as the second component (Fig.~\ref{fig:cnname}). 

\subsection{New {\revision $\mathsf{CN}$} Synchronization Process}
\label{subsec:sync}
When a new {\revision $\mathsf{CN}$} concludes the association process with a {\revision $\mathsf{CH}$}, the {\revision $\mathsf{CH}$} shares the new node information (such as its naming scheme) with its {\revision $\mathsf{CNs}$} and other {\revision $\mathsf{CH}$}s. Other {\revision $\mathsf{CH}$}s may further share this information with their own {\revision $\mathsf{CNs}$}. The rationale of sharing this information is to make {\revision other {\revision $\mathsf{CH}$} nodes and {\revision $\mathsf{CNs}$} } aware of the name of the new node, so that they can employ this name to fetch content in the future. Note that the synchronization process might be an expensive {\revision process --- its} cost grows with the number of nodes that participate in this process. {\revision Accordingly, the scope of the} synchronization process depends on the WSN application scenario and should be used according to the application requirements. {\revision The total number of nodes $N_{Rsync}$ receiving a sync message from {\revision an $h \in \mathsf{CHs}$ that} shares the information of a new node that joined its cluster is: %{\revision The number of transmissions $N_{Tsync}$ for an $h \in \mathsf{CH}$ to share the information of a new node with other nodes during this process is:

%\begin{equation} 
%     N_{Rsync} = N_{\mathsf{CN}}(h) + N_{\mathsf{CH}}(h) + \sum_{i=1}^{N_{\mathsf{CH}}(h)} a(i) * N_{\mathsf{CH}}(i)
%     \label{eq:sync}
%\end{equation}

\begin{equation} 
     N_{Rsync} = N_{\mathsf{CN}}(h) + N_{\mathsf{CH}}(h) + \sum_{i=1}^{N_{\mathsf{CH}}(h)} a(i) * N_{\mathsf{CN}}(i)
     \label{eq:sync}
\end{equation}

%\begin{equation} 
%     N_{Tsync} = 1 + \sum_{i=1}^{N_{\mathsf{CH}}(h)} a(i)
%     \label{eq:sync}
%\end{equation}

\noindent where $N_{\mathsf{CN}}(h)$ is the number of child nodes of $\mathsf{CH}$ $h$ that initiated the sync process and $N_{\mathsf{CH}}(h)$ is the number of CH nodes that {\revision are in the communication range of $\mathsf{CH}$ $h$}, thus receiving $h$'s sync message. $N_{CN}(i)$ is the number of $\mathsf{CN}$s that a $\mathsf{CH}$ $i$ has, which received $h$'s sync message and $a(i) = 1$ if a $\mathsf{CH}$ $i$ shares the new node information with its own {\revision $\mathsf{CNs}$ } ($a(i) = 0$ otherwise).}

\begin{figure}[!t]
	\centering
	\includegraphics[scale=0.36]{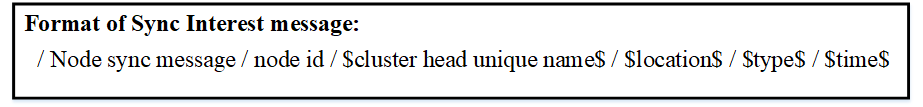}
	\caption{Synchronization naming scheme}
	\label{fig:syncname}
\end{figure}

Fig.~\ref{fig:syncname} shows the synchronization naming scheme that is used to share the new node information with other nodes in the network. The first name component (\texttt{Node\_Sync\_Message}) indicates that this is a synchronization packet. The remaining name components are the same as defined in the previous subsections and demonstrate the information of a new node. As soon the node is associated with the {\revision $\mathsf{CH}$}, the {\revision $\mathsf{CH}$} generates a synchronization Interest under the naming scheme of Fig.~\ref{fig:syncname}. 
On receiving a synchronization Interest, the {\revision $\mathsf{CH}$} and {\revision $\mathsf{CN}$} nodes (if enabled) store the new node information in their members-collection data structures.

\subsection{$\mathsf{\textit{CN}}$ Mobility}

CCIC-WSN enables seamless mobility of {\revision CNs}. When a {\revision CN} moves from one CH to another CH, it re-initiates the join process. During the association process with a new CH, the previous CH of this CN may receive the Interest with the information of the new CH. As a result, the old CH will deduce that this {\revision CN}  is no longer part of its cluster. 
It will also remove the name of this child from its members-collection data structure.

If a {\revision CN}  has moved out of the communication range of its old CH, the information or association packets with a new CH will not be received by the old CH. In this case, when the association process is completed, the new CH will share the information with the old CH via the synchronization process. As a result, the old CH will deduce that the {\revision CN}  has moved out of the communication range of the old cluster and now produces data as a part of a new cluster. 

{\revision After initializing the association process with a new CH, the child node's old CH will be notified that the {\revision CN}  has moved after a time interval $t_{disassociation}$ = $t_{join}$ + $t_{sync\_new CH->oldCH}$, where $t_{join}$ is the time needed for the {\revision CN}  to join the new CH (defined in Eq.~\ref{eq:join}) and $t_{sync\_new CH->oldCH}$ is the time needed for the {\revision CN}  information to be propagated from the new CH to the old CH through the synchronization process. 

}
\subsection{  CCIC-WSN: Forwarding Scenarios}
In cluster-based WSNs, the communication can be either intra- or inter-cluster. The intra-cluster communication can be further divided into three subcategories: (i) {\revision $\mathsf{CH}$} initiated, pull-based communication; (ii) {\revision $\mathsf{CN}$ initiated, pull-based communication}; and (iii) {\revision $\mathsf{CN}$} initiated, push-based communication. Similarly, the inter-cluster communication can also be further divided into three subcategories: (i) {\revision $\mathsf{CN}$} node to another cluster {\revision $\mathsf{CN}$} {\revision forwarding,} (ii) {\revision $\mathsf{CN}$} to another cluster {\revision $\mathsf{CH}$} {\revision forwarding,} and (iii) {\revision $\mathsf{CH}$} to {\revision $\mathsf{CH}$} forwarding. In the rest of this subsection, we describe the forwarding of Interest and Data packets in these intra- and inter-cluster communication cases.
\begin{figure*}
	\centering
	\includegraphics[width=1.0\linewidth]{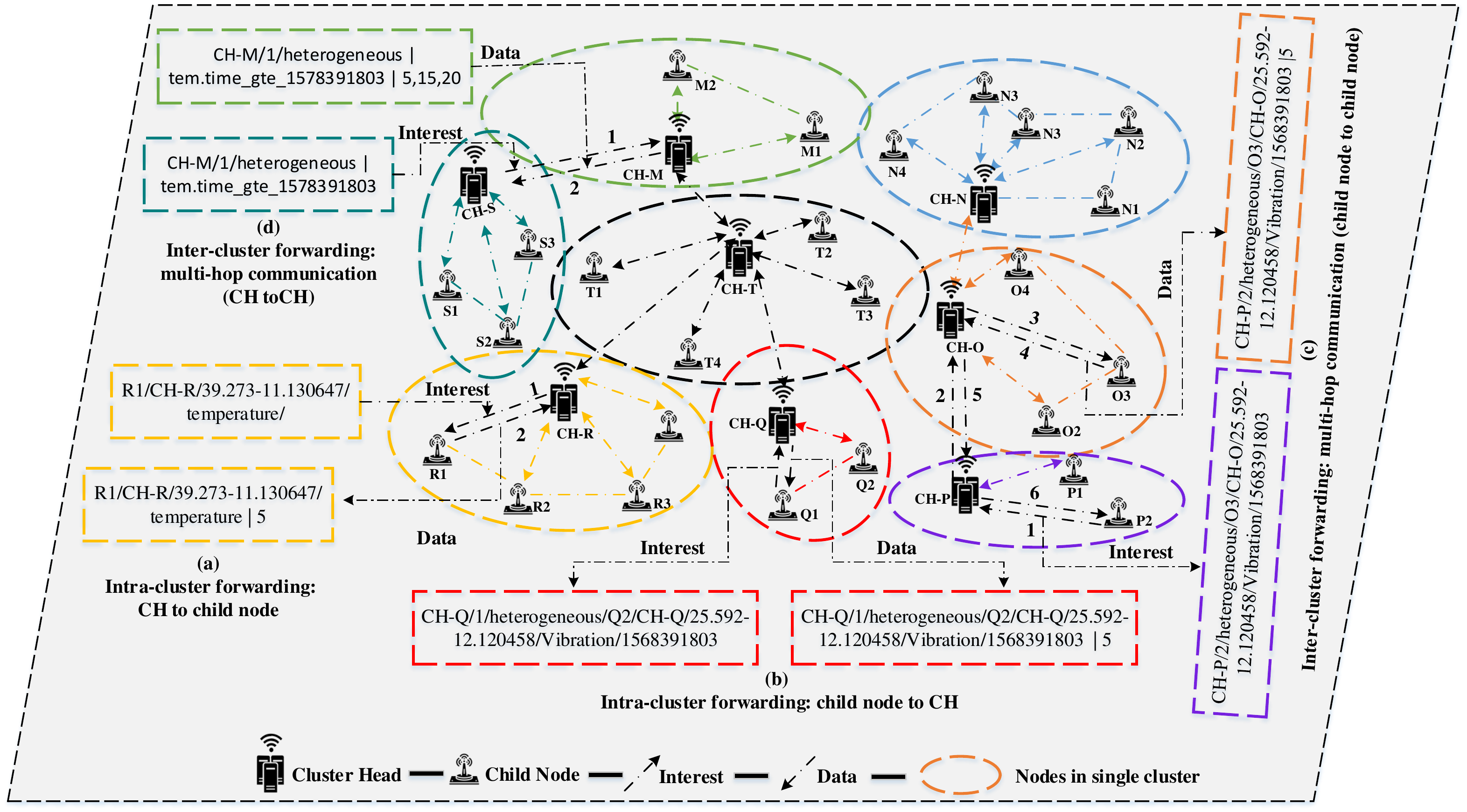}
	\caption{{\revision CCIC-WSN Forwarding Scenarios}}
	\label{fig:CCICForwarding}
\end{figure*}
\subsubsection{Intra-cluster forwarding: {\revision $\mathsf{\textit{CH}}$} to {\revision $\mathsf{\textit{CN}}$}}
\label{subsec:ch-to-cn}
In this case, {\revision $\mathsf{CH}$} acts as a consumer and can issue an Interest packet that carries the name for content to be retrieved from a {\revision $\mathsf{CN}$} one hop away. {\revision As all {\revision $\mathsf{CNs}$} are one hop} away from their corresponding {\revision $\mathsf{CH}$}, only the {\revision $\mathsf{CN}$} that produced the data responds, while other nodes in the cluster ignore the Interest packet. 

{\revision The use-case scenario (a) in Fig.~\ref{fig:CCICForwarding} illustrates a communication scenario of} Intra-cluster forwarding from a {\revision $\mathsf{CH}$} to a {\revision $\mathsf{CN}$}. The CH generates an Interest with a name “\texttt{R1/CH-R/39.273-11.130647/temperature/}”, where “R1” is the prefix of the {\revision $\mathsf{CN}$} that has produced the data, “CH-R” is the name prefix of the {\revision $\mathsf{CH}$}, “39.273-11.130647” refers to the location of the {\revision $\mathsf{CN}$}, and “temperature” shows the type of the requested data. When CN-R1 receives the Interest, it returns its temperature values to the CH-R. For example, "\texttt{R1/CH-R/39.273-11.130647/temperature/\\1578391803 \textbar 5}” indicates that a temperature value “5” is sent to CH-R by CN-R1, which is located at 39.273-11.130647 and sensed the value at time 1578391803. 

\subsubsection{Intra-cluster forwarding: {\revision $\mathsf{\textit{CN}}$} to {\revision $\mathsf{\textit{CH}}$}}
\label{subsec:intra}
In this case, a {\revision $\mathsf{CN}$} acts as a consumer and a {\revision $\mathsf{CH}$} acts as a producer. When a {\revision $\mathsf{CH}$} receives an Interest packet, it replies with a Data packet, while all other nodes within the {\revision $\mathsf{CH}$}'s communication range remain silent. {\revision The use-case scenario (b) in Fig.~\ref{fig:CCICForwarding} shows} a communication scenario of intra-cluster forwarding from a {\revision $\mathsf{CN}$} to a {\revision $\mathsf{CH}$}. {\revision A CN-Q1 generates an Interest packet with the name} “CH-Q/1/heterogeneous/Q2/CH-Q/25.592-12.120458/Vibration/1568391803“, where “CH-Q“ refers to the name of the CH, “1“ shows that CH-Q is one hop away from the sink node, “Q2“ refers to the name of the $\mathsf{CN}$, “25.592-12.120458“ is the location of CN-Q2, “vibration“ is the type of the requested data, and “1568391803“ is the epoch time when the requested data was produced\footnote{We assume that nodes have synchronized clocks. This can be achieved, for example, through a GPS system or a network-based solution~\cite{mtibaa2020ndntp}. If the epoch time in an Interest does not match the epoch time of the actual data production (e.g., due to clock drifting), the CH sends back the data that has been generated as close to the requested epoch time as possible.}. When CH-Q receives the Interest, it either {\revision retrieves the content from CN-Q2 by following the CH-to-CN (Section ~\ref{subsec:ch-to-cn}) }forwarding mechanism or returns the stored {\revision value,} if available.

\subsubsection{Intra-cluster forwarding (push scenario): {\revision $\mathsf{\textit{CN}}$} to {\revision $\mathsf{\textit{CH}}$}}
This mechanism is designed for a push-based communication scenario, where a child node may send a {\revision Data} packet directly to a CH. In emergency scenarios (e.g., detection of an individual with an infectious diseases, car accidents), {\revision data can be attached to Interests.} These Interests will be broadcast from a child node to a CH. The CH will respond with a Data packet to acknowledge the reception of this Interest, while other child nodes that receive this Interest will ignore it. The names of the exchanged packets follow the naming scheme presented in Section~\ref{subsec:intra}.

\subsubsection{Inter-cluster forwarding: multi-hop communication ({\revision $\mathsf{\textit{CN}}$} to {\revision $\mathsf{\textit{CN}}$})}
\label{subsec:inter}
This mechanism is used for inter-cluster communication scenarios, where a {\revision $\mathsf{CN}$} may fetch the content from another {\revision $\mathsf{CN}$} that is a {\revision member} of another cluster. In such cases, only the CHs of the consumer and producer child nodes will participate in the communication and act as forwarders. We explain the 
Interest and Data exchange mechanism in a stepwise manner through an example {\revision presented as use-case scenario (c) in Fig.~\ref{fig:CCICForwarding} as follows:}

\textit{Step 1:} A {\revision CN-P2, which is a member of CH-P, is interested in the content of CN-O3, which is a member of CH-O. Therefore, the CN-P2 sends an Interest to request the content produced by CN-O3 of CH-O. In this case, the CN-P2 is the consumer, CH-P and CH-O are forwarders, while the CN-O3 is the producer}\footnote{Note that for communication with child nodes in another cluster, we assume that the required information has been shared during the synchronization process (Section~\ref{subsec:sync}). }.    

\textit{Step 2:} When {\revision the} CH-P receives the Interest from {\revision CN-P2}, it broadcasts this Interest within its communication range. All other child nodes associated with CH-P discard the Interest. 

\textit{Step 3:} When the CH-O receives the Interest from CH-P, it broadcasts this Interest within its communication range. All the nodes will discard the Interest except for {\revision CN-O3} that is the producer of the requested data. 

\textit{Step 4:} {\revision CN-O3} receives the Interest from CH-O and either checks whether the content has already been generated or senses the environment in real-time (if a timestamp in the future is specified in the Interest). It then responds with a Data packet containing the requested content.

\textit{Step 5:} CH-O receives the Data packet from {\revision CN-O3}. It may cache the data in its CS and then forward it to CH-P. Due to their resource-rich nature, CHs may cache received Data packets in their CS. 

\textit{Step 6:} CH-P receives the Data packet from CH-O and forwards it to child node P2 after potentially caching it in its CS.

\subsubsection{Inter-cluster forwarding: multi-hop communication ({\revision $\mathsf{\textit{CN}}$} to {\revision $\mathsf{\textit{CH}}$})}
This mechanism is used by a child node of a cluster in order to retrieve some content from the CH of another cluster. In the {\revision above} example of Fig.~\ref{fig:CCICForwarding} use-case scenario (c), the {\revision CN-P2} sent an Interest to retrieve some content from CH-O. The request follows similar steps as mentioned in Section~\ref{subsec:inter} with the difference that, in current use-case scenario, {\revision CH-O} will not forward the Interest to one of its child nodes, but it will rather directly respond the Interest packet if the requested content is available in its CS. To avoid unnecessary packet transmissions in the network, only {\revision CN-P2}, CH-P itself, and CH-O participate in the communication process. 

\subsubsection{Inter-cluster forwarding: multi-hop communication ($\mathsf{\textit{CH}}$ to $\mathsf{\textit{CH}}$)}
In this scenario, a CH communicates with another CH to retrieve some content or request data based on lite queries. For instance, data aggregation operations may be needed by a CH. In such cases, the CH sends Interest packets to another CH to receive aggregated data.  
{\revision The use-case scenario (d) in Fig.~\ref{fig:CCICForwarding} illustrates a scenario,} where CH-S sends an Interest with a lite-query in its name to retrieve a number of temperature values recorded after a certain point in time. In our example, the epoch time “1578391803” indicates the time of January 7, 2020, 10:10:03 AM. As a result, this Interest will retrieve the temperature values from CH-M that have been generated after this point in time.

\section{Evaluation}
\label{sec:evaluation}
In this section, our goal is to evaluate the tradeoffs of the CCIC-WSN design through an extensive simulation study and compare its performance with state-of-the-art schemes for WSNs; {\revision Energy-Efficient} Interest Forwarding (EEIF)~\cite{EEIF} and Hierarchical and Flat-based Hybrid Naming scheme (HFHN)~\cite{HFHN}, {\revision both of which have been} discussed in Section~\ref{subsec:relWSN}. As our baseline, we also compare CCIC-WSN to the {\revision vanilla} NDN architecture without introducing any modifications for WSNs. We first present our simulation setup and we then present our simulation results.

\subsection{Simulation Setup}
\begin{figure}[!t]
	\centering
	\includegraphics[scale=0.45]{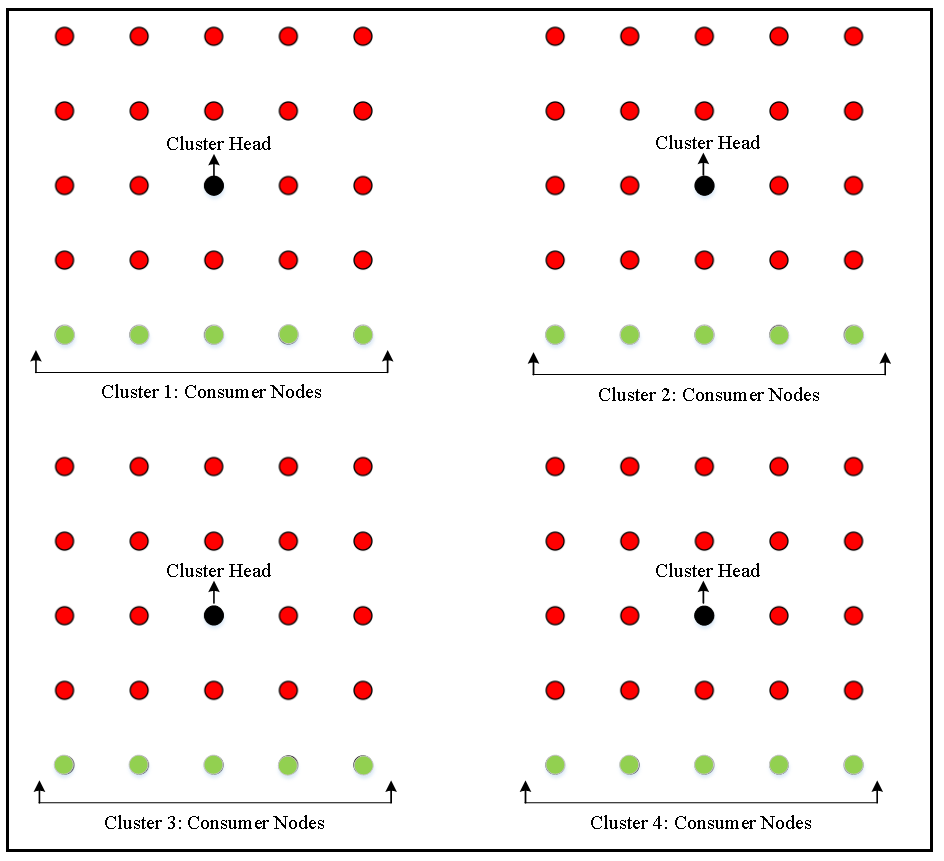}
	\caption{Simulation Topology}
	\label{fig:simulation}
	\vspace{-0.5cm}
\end{figure}

In our simulations, we consider multiple nodes that are deployed in the form of clusters in an area of 300 m $\times$ 200 m with a total of 100 nodes (Fig.~\ref{fig:simulation}). Each cluster consists of one {\revision $\mathsf{CH}$} node and multiple {\revision $\mathsf{CNs}$}. For our simulations, we implemented CCIC-WSN in ndnSIM~\cite{ndnSIM}, the NDN simulator, on a computer equipped with a Core i7 CPU and 8GB of RAM running Linux Ubuntu. The transmission range of {\revision $\mathsf{CNs}$} is limited to one hop (i.e., to the {\revision $\mathsf{CH}$} they are associated with), while the transmission range of $\mathsf{CH}$s is higher and covers the entire cluster. The $\mathsf{CH}$ is deliberately located at the center of a cluster and is reachable by its $\mathsf{CNs}$. In our simulation topology (Fig.~\ref{fig:simulation}), the green nodes at the bottom of a cluster are randomly chosen as the consumer nodes that request content at different rates. The black nodes are chosen as the CH, while the remaining nodes are producers. A summary of our simulation parameters is presented in Table~\ref{table:setup}.

In our simulation study, we considered the following evaluation metrics: 

\begin{itemize}[leftmargin=*]
	\item \textbf{Energy consumption:} The total energy ({\revision ($e$)} in J) consumed by all the nodes {\revision ($N$)} for the transmission of Interest and Data packets. 
	
	{\revision $$
	\text{Energy~Consumption} = \sum_{i = 1}^{N} e_i^{I_{pkts}} +  e_i^{D_{pkts}}
	$$ }
	
 	\item \textbf{Interest satisfaction rate:} The ratio of the total number of Data packets {\revision ($D_{pkt}$)} successfully retrieved  to the total number of Interest packets {\revision ($I_{pkt}$)} generated in the network.
	{\revision $$
	\text{Interest~satisfaction~rate} = \frac{\sum_{i = 1}^{N} D_{pkt}^{i}}{\sum_{i = 1}^{n} I_{pkt}^{i}}
	$$ }
	
	\item \textbf{Query satisfaction rate:} The ratio of the total query-based Interests satisfied {\revision ($Q_{I\prime}$)}  to the total number of query-based Interests generated {\revision $Q_{I}$}. 

    {\revision{} $$
	\text{Interest~satisfaction~rate} = \frac{\sum_{i = 1}^{N} Q_{I\prime}^{i}}{\sum_{i = 1}^{n} Q_{I}^{i}}
	$$ }
	
	\item \textbf{Interest satisfaction delay:} The sum of the time required for an Interest {\revision $I$} to reach the producer $p$,  {\revision ($t_{I, p}$)}, for the producer to process {\revision (${P\prime}$)} this Interest {\revision ($t_{I,p}^{P\prime}$)}, and for the corresponding Data packet to reach the consumer $c$, {\revision ($t_{D, c}$)}. 
	{\revision $$
	\text{Interest~satisfaction~delay} = t_{I, p} + t_{I,p}^{P\prime} + t_{D, c}
	$$ }
	
	\item \textbf{Query satisfaction delay:} The sum of the time required for an Interest carrying a lite-query {\revision $Q$} to reach the producer {\revision $p$}, {\revision ($t_{Q, p}$)}, for the producer to process {\revision (${P\prime}$)} this lite-query Interest {\revision ($t_{Q,p}^{P\prime}$)}, and for the corresponding Data packet to reach the consumer {\revision $c$,  ~($t_{D, c}$)}. 
	{\revision $$
	\text{Query~satisfaction~delay} = t_{Q, p} + t_{Q,p}^{{P\prime}} + t_{D, c}
	$$ }
	
	\item \textbf{Node association time:} The time required for the association {\revision ($t_{association}$)} of a CN node with a CH.
	{\revision $$
	\text{Node~association~time} = \frac{\sum_{i = 1}^{|CN|} t_{association}^{i}}{|CN|}
	$$ }

	\item \textbf{New node sync time:} The time required {\revision ($T_{sync}$)} for a CH to share the information of a newly associated child node with its other child nodes and other CHs {\revision ($T_{sync}^{CH}$)}, and these CHs to share this information with their own child nodes {\revision ($T_{sync}^{CNs}$)} through the synchronization process.
	{\revision $$
	{\revision T_{sync}} = \frac{\sum_{i = 1}^{N} T_{sync}^{CH, i}}{N} + \frac{\sum_{j=1}^{M} a(j) * T_{sync}^{CNs, j}}{M}
	$$ }

\end{itemize}

\subsection{Evaluation Results}

\subsubsection{Energy Consumption}
\begin{figure}[!t]
	\centering
	\includegraphics[scale=0.34]{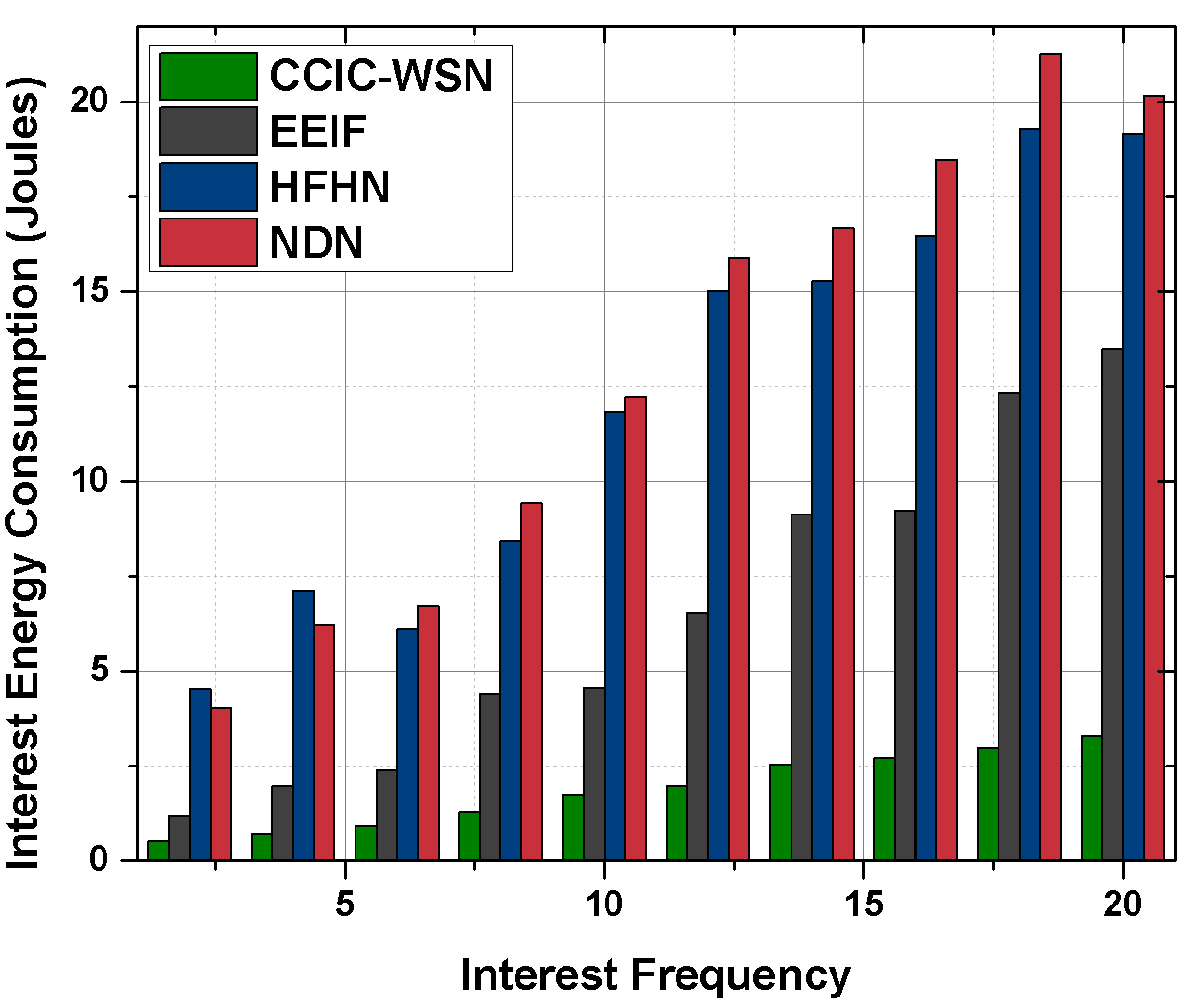}
	\caption{Interest energy consumption as a function of Interest
frequency}
	\label{fig:interestenergy}
	\vspace{-0.5cm}
\end{figure}

\begin{figure}[!t]
	\centering
	\includegraphics[scale=0.34]{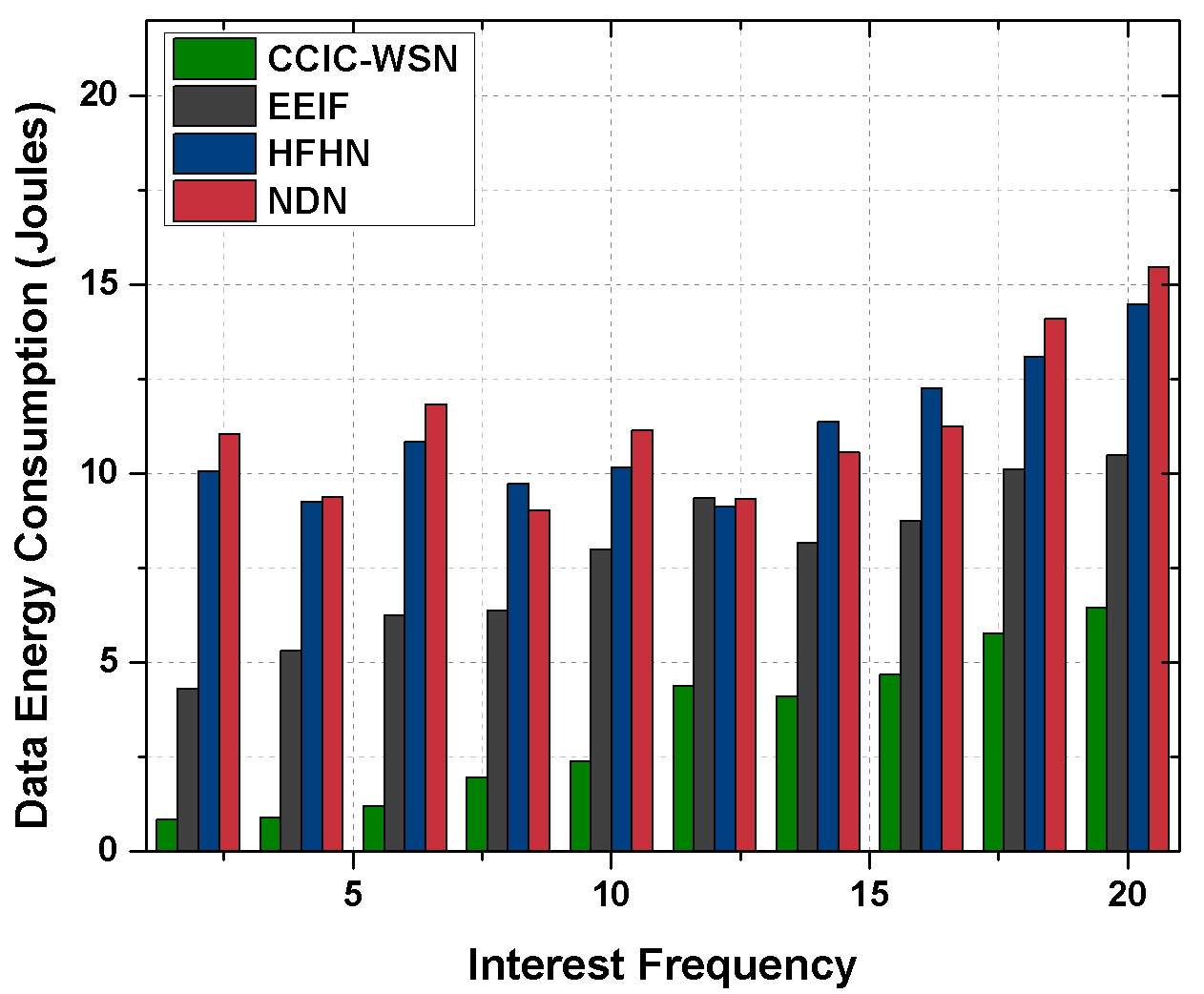}
	\caption{Data energy consumption as a function of Interest
frequency}
	\label{fig:dataenergy}
	\vspace{-0.5cm}
\end{figure}

Energy consumption is directly proportional to the number of Interest and Data packets transmitted in the network. Fig.~\ref{fig:interestenergy} and  Fig.~\ref{fig:dataenergy} {\revision demonstrate a comparative analysis of Interest and Data packets energy consumption among vanilla} NDN, HFHN, EEIF, and CCIC-WSN. For this experiment, we vary the number of the Interest generation frequency from 2 Interest/sec to 20 Interest/sec and calculate the total amount of energy consumed by both Interest and Data packet transmissions.  

\begin{table}[!t]
	\caption{Simulation Parameters}
	\begin{tabular}{|c|c|}
		\hline
		\textbf{Parameter}                       & \textbf{Value}                                                                                                        \\ \hline
		\multicolumn{1}{|m{3cm}|}{Simulator} & 
		\multicolumn{1}{m{5cm}|}{NS-3 (ndnSIM 2.5)} 
		\\ \hline
		\multicolumn{1}{|m{3cm}|}{Communication Stack} & 
		\multicolumn{1}{m{5cm}|}{NDN} 
		\\ \hline
		\multicolumn{1}{|m{3cm}|}{Wireless Interface} & 
		\multicolumn{1}{m{5cm}|}{IEEE 802.15.4} 
		\\ \hline
		\multicolumn{1}{|m{3cm}|}{Topology size} & 
		\multicolumn{1}{m{5cm}|}{\text{300$\times$200}} 
		\\ \hline
		\multicolumn{1}{|m{3cm}|}{Total Number of Nodes} & 
		\multicolumn{1}{m{5cm}|}{100} 
		\\ \hline
		\multicolumn{1}{|m{3cm}|}{Total Number of Clusters} & 
		\multicolumn{1}{m{5cm}|}{4} 
		\\ \hline
		\multicolumn{1}{|m{3cm}|}{Mobility Model} & 
		\multicolumn{1}{m{5cm}|}{Constant Mobility with fixed location} 
		\\ \hline
		\multicolumn{1}{|m{3cm}|}{Propagation Delay Model} & 
		\multicolumn{1}{m{5cm}|}{ConstantSpeedPropagationDelayModel} 
		\\ \hline
		\multicolumn{1}{|m{3cm}|}{Energy consumption per bit} & 
		\multicolumn{1}{m{5cm}|}{0.5 $\mu$J/bit } 
		\\ \hline
		\multicolumn{1}{|m{3cm}|}{PIT Timer} & 
		\multicolumn{1}{m{5cm}|}{4 sec} 
		\\ \hline
		\multicolumn{1}{|m{3cm}|}{Interest Packet Size} & 
		\multicolumn{1}{m{5cm}|}{48 bytes} 
		\\ \hline
		\multicolumn{1}{|m{3cm}|}{Data Packet Size} & 
		\multicolumn{1}{m{5cm}|}{96 bytes} 
		\\ \hline
		\multicolumn{1}{|m{3cm}|}{Simulation Time} & 
		\multicolumn{1}{m{5cm}|}{1800s} 
		\\ \hline
		
	\end{tabular}%
	\label{table:setup}
	\vspace{-0.2cm}
\end{table}

{\revision Analyzing Fig.~\ref{fig:interestenergy} and  Fig.~\ref{fig:dataenergy} for vanilla NDN,} our results indicate that for {\revision low rates of Interest generation,} the energy consumption of Interest packets is lower compared to the energy consumption of Data packets at identical frequency rate, since the {\revision size of an Interest packet} is almost half the size of {\revision a} Data packet. As the Interest generation frequency increases, the Interest packet energy consumption also increases and surpasses the Data packet energy consumption. This is due to the ad hoc nature of communication among the nodes, where each node that receives an Interest, forwards the Interest to neighboring nodes in the absence of the requested data. This creates a broadcast storm in the cluster and collisions occur due to simultaneous transmissions. {\revision Considering} such collisions occur frequently, they increase the Interest retransmissions and, subsequently, the Interest energy consumption exceeds the Data packet energy consumption. 


{\revision Similar to vanilla NDN,} HFHN demonstrates an analogous trend in results for Interest and Data packets energy consumption. The reason for this behavior is that in a single cluster all the nodes act as forwarders; {\revision rather than discarding the received Interests, they forward them instead,} creating an Interest storm in the cluster. {\revision Contrarily, the results for EEIF} show that it controls Interest flooding through scope control and packet suppression. At the same time, EEIF forwards Interests in dual mode (controlled flooding or directive mode). {\revision Thus,} EEIF selects a limited number of forwarding nodes for the transmission of Interest and Data packets. However, our results indicate that 8 to 12 nodes out of the 25 nodes of a cluster still participate in {\revision the communication.} 
\begin{figure*}
	
	\begin{minipage}{\columnwidth}
		\includegraphics[width=\linewidth,height=0.27\textheight]{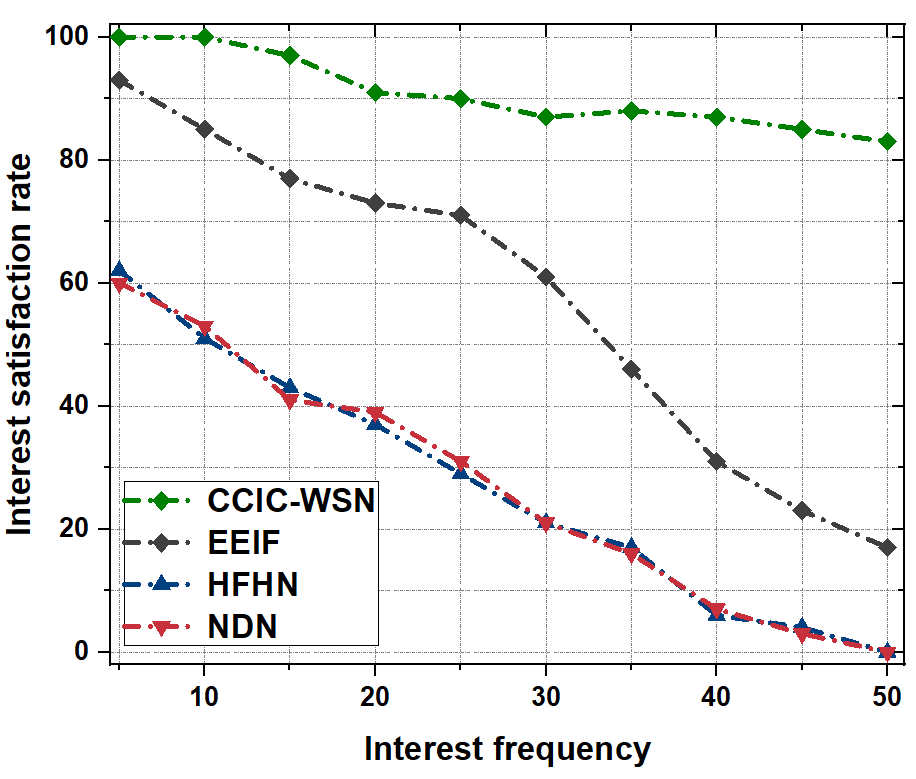}
		\caption{Interest satisfaction rate as a function of the Interest generation frequency}
		\label{fig:isr}
	\end{minipage}
	\hfill  % maximize the space between the minipages
	\begin{minipage}{\columnwidth}
		\includegraphics[width=\linewidth,height=0.27\textheight]{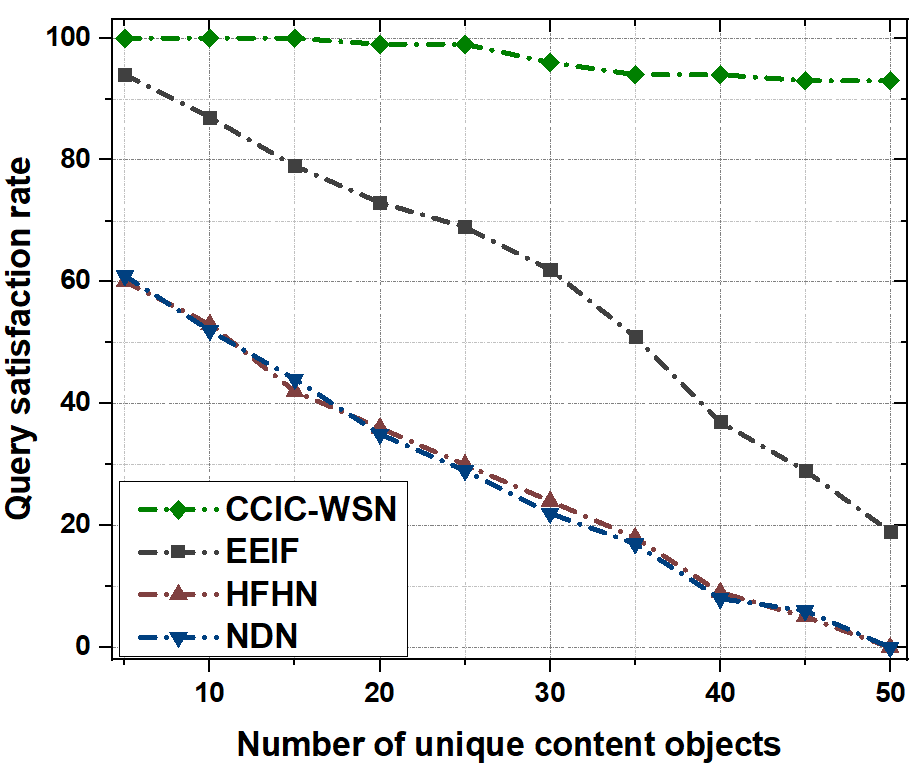}
		\caption{Query satisfaction rate as a function of the number of unique content objects}
		\label{fig:qsr}
	\end{minipage}
	\vspace{0.5cm}
\end{figure*}
\begin{figure*}
	\vspace{-0.5cm}
	\begin{minipage}{\columnwidth}
		\includegraphics[width=\linewidth,height=0.27\textheight]{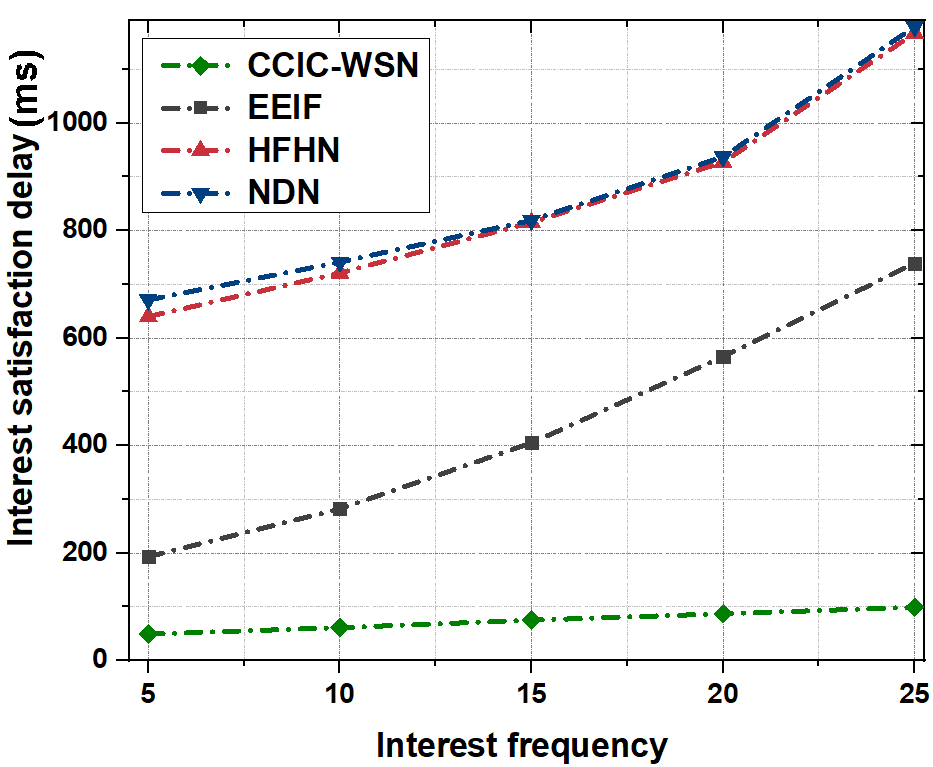}
		\caption{ Interest satisfaction delay as a function of Interest frequency }
		\label{fig:isd}
	\end{minipage}
	\hfill  % maximize the space between the minipages
	\begin{minipage}{\columnwidth}
		\includegraphics[width=\linewidth,height=0.27\textheight]{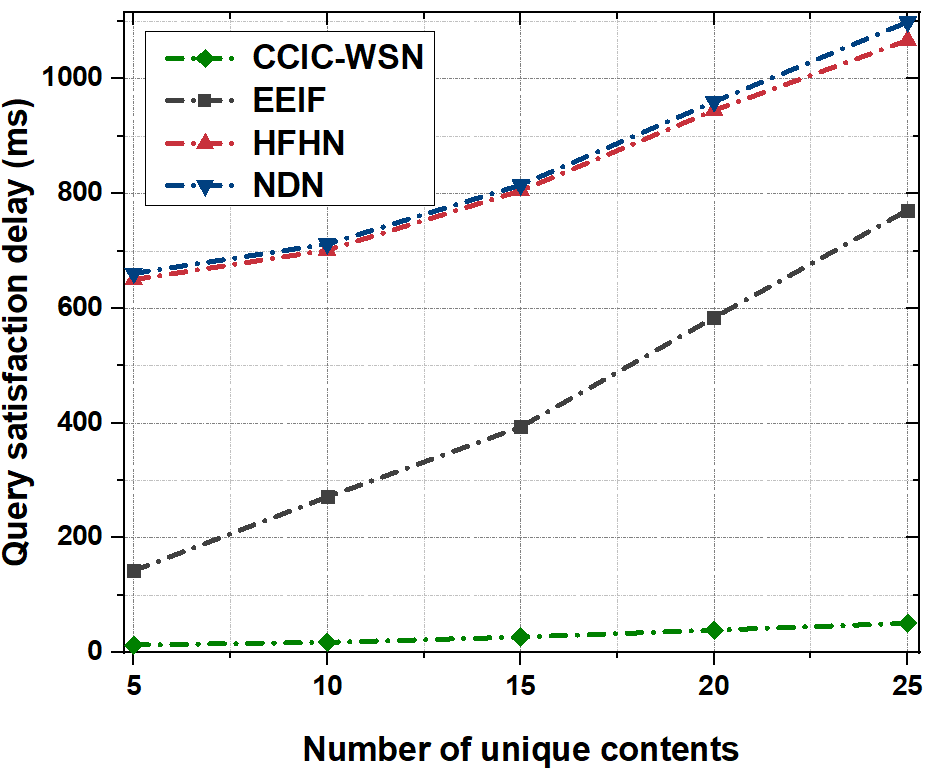}
		\caption{ Query satisfaction delay as a function of number of unique content }
		\label{fig:qsd}
	\end{minipage}
	\vspace{-0.5cm}
\end{figure*}
{\revision In comparison, our findings} show that the use of CCIC-WSN results in 71–90\% lower total consumed energy compared to vanilla NDN, HFHN, and EEIF. For intra-cluster communication, nodes in CCIC-WSN communicate directly with producer nodes or CHs without the need {\revision for} any intermediate forwarders. As a result, CCIC-WSN can effectively control the transmissions of Interest and Data packets, avoiding storms of broadcast packets, {\revision and} minimize collisions in the network, subsequently minimizing the overall consumed energy across the network.

\subsubsection{Interest and Query Satisfaction Rates}
Fig.~\ref{fig:isr} and Fig.~\ref{fig:qsr} show results on the Interest Satisfaction Rate (ISR) and the Query Satisfaction Rate (QSR){\revision ,} respectively.  
For ISR, we vary the Interest generation frequency between 10 Interest/sec and 50 Interest/sec, while, for QSR, we vary the number of unique content objects between 10 and 50 objects. For QSR, we assumed that producers have pre-recorded data samples{\revision ,} and all of them are uniquely identified by a time value in their names.

The results presented in Fig.~\ref{fig:isr} demonstrate that CCIC-WSN performs better than the other three schemes. The reason is that CCIC-WSN directly fetches the content from the producer for intra-cluster communication and  in the worst case scenario, it involves two forwarder nodes for inter-cluster communication. {\revision In contrast,} the remaining three schemes may involve multiple intermediate forwarders in the communication, causing congestion and collisions in the network and {\revision thus} decreasing ISR.

\begin{figure*}
	\begin{minipage}{\columnwidth}
		\includegraphics[width=\linewidth,height=0.27\textheight]{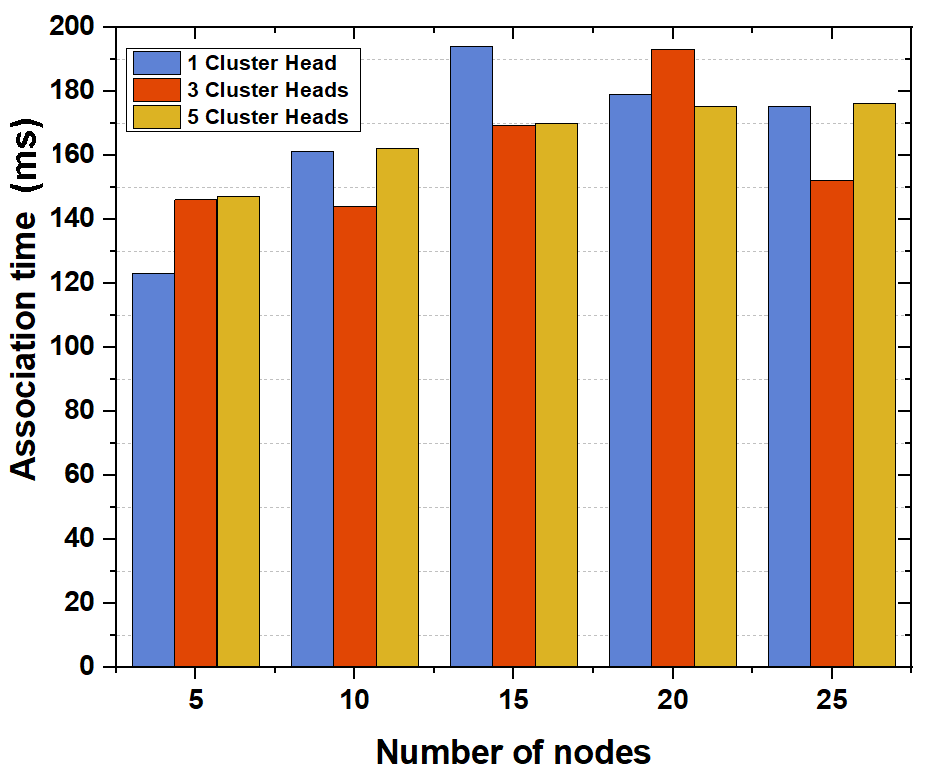}
		\caption{Node association time as a function of the number of nodes in the cluster and the number of CHs within the communication range of the new node}
		\label{fig:assoctime}
	\end{minipage}
	\hfill  % maximize the space between the minipages
	\begin{minipage}{\columnwidth}
		\includegraphics[width=\linewidth,height=0.27\textheight]{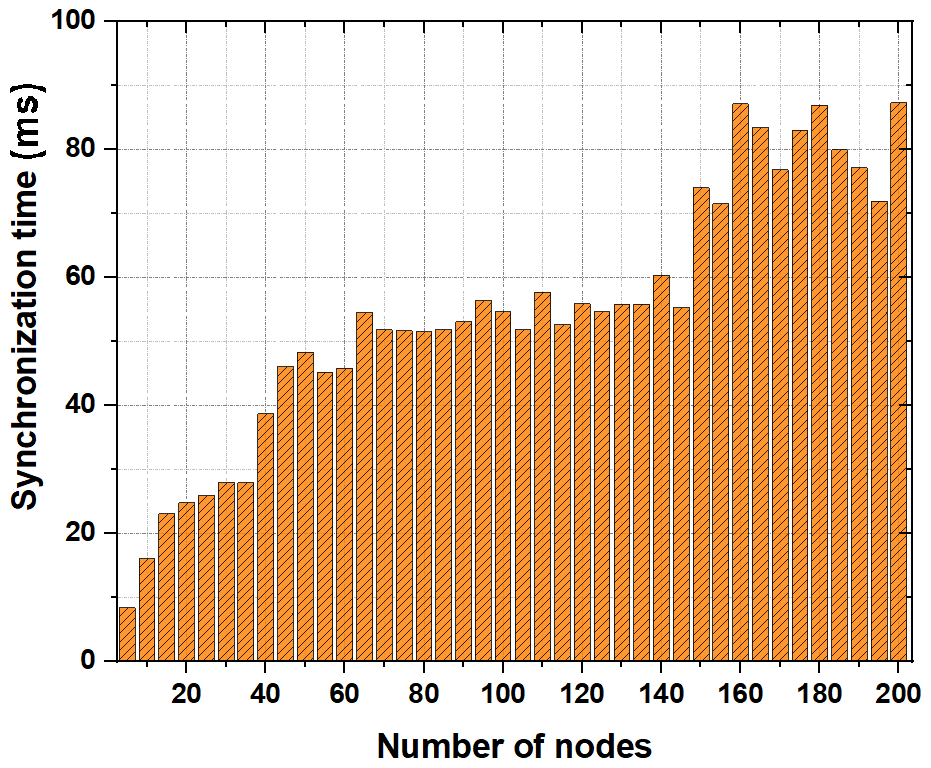}
		\caption{Synchronization time as a function of the total number of nodes}
		\label{fig:synctime}
	\end{minipage}
\end{figure*}
The results {\revision in} Fig.~\ref{fig:qsr} show that CCIC-WSN attains a considerably higher QSR {\revision compared to} EEIF, HFHN{\revision , and vanilla} NDN. A single lite-query-based Interest packet in CCIC-WSN can fetch multiple content objects (on average 4 to 5) in a single Data packet by specifying the exact logic in its lite-query component. {\revision Thus}, CCIC-WSN can decrease the number of required Interests for the retrieval of the same amount of content up to 4 times on average as compared to EEIF, HFHN{\revision , and vanilla} NDN. 

\subsubsection{Interest and Query Satisfaction Delays}
For the Interest Satisfaction Delay (ISD), {\revision  we vary the Interest generation frequency between 5 Interest/sec and 25 Interest/sec, as shown in Fig.~\ref{fig:isd}}, while for the Query Satisfaction Delay (QSD), the number of unique content objects varies between 5 and 25 objects{\revision , as shown in Fig.~\ref{fig:qsd}}.  
Our results show that ISD for CCIC-WSN is significantly lower than the other three schemes. Furthermore, CCIC-WSN significantly reduces QSD compared to the other three schemes; a single lite-query-based Interest fetches several content objects as a single Data packet, reducing the delay needed to fetch the same amount of data compared to sending multiple and separate Interests for each content object. 
Overall, our results demonstrate that CCIC-WSN achieves 74-96\% lower ISD and QSD than EEIF, HFHN{\revision , and vanillain NDN}.

\subsubsection{Node Association Time}
In these experiments, we assume that new nodes join the network during the simulation execution. In such cases, existing CHs and already associated child nodes exchange Interest and Data packets with each other. {\revision We present results in Fig.~\ref{fig:assoctime} on} the association time as a function of the number of nodes already in a cluster (5 to 25 nodes) and the number of CHs that are within the communication range of the new node (1 to 3 CHs). The results demonstrate that the maximum association time for a node is approximately 194ms. Moreover, the results show that the association time does not alter considerably as we increase the number of CHs within the communication range of the new node, {\revision because} only CHs that have available resources to accommodate the new node will reply to the node's join request. {\revision As a result,} the new node will be able to seamlessly and quickly associate itself with a CH that has available resources.

\subsubsection{New Node Sync Time}
In these experiments, we assume that the synchronization process happens once a new node associates with a CH. We measured the synchronization time by varying the total number of nodes in the network from 20 to 200{\revision, as shown in Fig~\ref{fig:synctime}}. The results show that the synchronization time increases with the number of nodes in the network. The maximum recorded time is 87 ms when the total number nodes in the network is 200.

\section{Future Directions and Open Research Challenges}
\label{sec:future}
The CCIC-WSN architecture fulfills the fundamental and necessary requirements of cluster-based WSN. However, this is only the first step {\revision toward} a concrete NDN-based architecture for cluster-based WSNs with several important areas yet to be explored.

\subsection{TDMA-based Slot Assignments}
To avoid interference and collisions, the child nodes in a single-channel cluster-based WSN forward sensed data in unique time slots. The CH splits the channel into equal time slots and distributes these time slots to the child nodes. The CH also communicates synchronization information to child nodes throughout the cluster. The CH can share such synchronization information in a control time slot which should be at the start of the time window. We may call these time slots as member-slots. In addition to the member-slots, a CH slot is also needed, which can be further divided into sub-slots for the association of new nodes with the CH and the synchronization with other CHs. To support such TDMA-based operations, further research is needed to explore energy-efficient slot assignment {\revision mechanisms} and naming schemes under which the slot information will be shared.

\subsection{Cluster Head Reshuffling} 
{\revision The CH consumes more energy compared to the CN, as most of the time the CH node performs compute-intensive tasks and forwards the Data packets over long distances \cite{energysurvey,energyfuzzy}. Consequently, after a few rounds, the elected CH may not be able to perform its duties and perish due to high energy consumption. To avoid communication losses in such failure scenarios, the expiring CH should handover its responsibilities to the nearby resourceful node. The CH handover process, often referred to as cluster-head reshuffling, may have an adverse effect on the performance of a cluster-based WSN, and if not carefully handled, it may result in high energy consumption and communication delays. To enable efficient CH reshuffling with the aim of minimizing the energy consumption, separate naming schemes are required for: 1) multi-casting of the expiration message among neighboring child nodes, 2) broadcasting of new CH name in the network, and 3) exchange of information between the old and the new CH. In addition, the proposed handover algorithm ought to demonstrate the minimum complexity and time requirement, and thus does not affect the overall communication requirements of cluster-based WSN.}

\subsection{Secure Association, Data Authentication, and Encryption}
\label{subsec:authentication}
Given that asymmetric cryptography may not be feasible for WSNs and IoT, since it may be too expensive for such applications~\cite{shi2004designing}, solutions based on symmetric cryptography need to be explored. For example, child nodes can authenticate the CH during the association process (and vice versa) based on pre-shared symmetric keys and a trusted resourceful node that acts as an authentication manager~\cite{mick2017laser, bersani2007eap}. Symmetric keys for Data packet authentication and encryption/decryption may be generated once a child node becomes securely associated with a CH. The used symmetric keys will need to be refreshed/rotated over time to minimize the potential exposure in cases {\revision where there is leakage of a symmetric key}~\cite{li2019secure}. Further research is needed along the direction of secure sensor and IoT on boarding, data encryption/decryption and authentication, {\revision and} to investigate the tradeoffs of the key rotation period.

\subsection{Query Execution Plans}
To improve query execution, modern databases such as MongoDB, MySQL, and SQL use execution plans that employ a query optimizer~\cite{kabra1998efficient, scherb2019execution}. In the proposed CCIC-WSN lite-query structure, a lite-query execution plan could be a set of sequences that a CH node performs to execute the lite-query. {\revision The} need to devise a proper query plan for CCIC-WSN ascends, because consumers in CCIC-WSN formulate their lite-queries through filtering logic, but they do not share the exact order for the query execution. 
{\revision Therefore}, a proper query planner module for CCIC-WSN should be developed. The responsibility of this type of module will be to fetch the best execution plan to efficiently filter the child nodes’ content. 

\subsection{Denial-of-Service Attack during New Node Association}
During the association process, a new node may receive the information of malicious CHs within its communication range. Malicious CHs may ignore the association requests of child nodes, essentially launching a Denial-of-Service (DoS) attack against these nodes. As a result, child nodes may not be able to associate with a legitimate CH in order to communicate with other nodes in the network.

\subsection{CCIC-WSN Cache Replacement Strategies}
Since Data packets in CCIC-WSN may contain content that is filtered by the logic defined in lite-query components, the lite-query should be stored in the CS of a CH along with the content. If the same lite-query arrives in the future, the result can be provided from the CS rather than from the original CH. In addition to that, further research is needed on cache replacement policies to investigate the impact and tradeoffs of such policies in WSNs~\cite{ullah2020icn}.

\section{Conclusion}
\label{sec:conclustion}
In this paper, we proposed CCIC-WSN, an NDN-based design for single-channel cluster-based WSNs. We first presented the motivation {\revision for} our work and then we discussed related work in ICN-based WSNs. We then presented the CCIC-WSN design components and evaluated CCIC-WSN through an extensive simulation study, where we compared its design to state-of-the-art solutions in ICN-based WSNs and a baseline {\revision vanilla} NDN communication scheme. Finally, we presented a number of research directions and open issues that we plan to investigate as part of our future research work.

\bibliographystyle{IEEEtran}
\bibliography{CCIC-WSN}

\vskip -2.5\baselineskip plus -1fil

\begin{IEEEbiographynophoto}{Muhammad Atif Ur Rehman}
	is currently pursuing a Ph.D. degree in Computer Engineering with the Broadband Convergence Networks Laboratory at Hongik University, South Korea. He received a B.S. degree in Electronics \& Communication from The University of Lahore, Lahore, Pakistan, in 2013, and an M.S. degree in Computer Science from COMSATS University, Islamabad, Pakistan, in 2016. His major interests are in the field of Information-Centric Wireless Networks, Named Data Networking, Wireless Sensor Networking, Edge Computing, Internet of Things, and 5th Generation communication. 
\end{IEEEbiographynophoto}

\vskip -2.5\baselineskip plus -1fil

\begin{IEEEbiographynophoto}{Rehmat Ullah}
is an Assistant Professor with the Department of Computer Engineering at Gachon University, South Korea. He received the B.S. and M.S. degrees in computer science from COMSATS University Islamabad, Pakistan, in 2013 and 2016, respectively, and the Ph.D. degree in electronics and computer engineering from Hongik University, South Korea, in February 2020. His research focuses on the broader area of future Internet and network systems, particularly the development of architectures, algorithms, and protocols for emerging paradigms, such as information-centric networking/named data networking (ICN/NDN), Internet of Things (IoT), cloud/edge/fog computing for IoT, and 5G and beyond. 
\end{IEEEbiographynophoto}

\vskip -2.5\baselineskip plus -1fil

\begin{IEEEbiographynophoto}{Byung-Seo Kim}
	(M'02-SM'17) is a Full Professor with the Department of Software and Communications Engineering, Hongik University, South Korea. He received his B.S. degree in Electrical Engineering from In-Ha University, In-Chon, Korea in 1998 and his M.S. and Ph.D. degrees in Electrical and Computer Engineering from the University of Florida in 2001 and 2004, respectively. His research interests include the design and development of efficient wireless/wired networks including link-adaptable/cross-layer-based protocols, multi-protocol structures, wireless CCNs/NDNs, Mobile Edge Computing, physical layer design for broadband PLC, and resource allocation algorithms for wireless networks.
\end{IEEEbiographynophoto}

\vskip -2.5\baselineskip plus -1fil

\begin{IEEEbiographynophoto}{Boubakr Nour} (GS'17, M'20) received his Ph.D. degree in Computer Science and Technology at the Beijing Institute of Technology, Beijing, China. His research interests include next-generation networking and the Internet. He is the recipient of the Best Paper Award at IEEE GLOBECOM (2018), and the Excellent Student Award at Beijing Institute of Technology in 2016, 2017, and 2018 consecutively.
\end{IEEEbiographynophoto}

\vskip -2.5\baselineskip plus -1fil

\begin{IEEEbiographynophoto}{Spyridon Mastorakis }
	is an Assistant Professor in Computer Science at the University of Nebraska, Omaha. He received his Ph.D. in Computer Science from the University of California, Los Angeles (UCLA) in 2019. He also received an MS in Computer Science from UCLA in 2017 and a 5-year diploma (equivalent to M.Eng.) in Electrical and Computer Engineering from the National Technical University of Athens (NTUA) in 2014. His research interests include network systems, Internet architectures and protocols, IoT and edge computing, and security.
\end{IEEEbiographynophoto}

\end{document}